\documentclass[11pt,a4paper]{article}
\usepackage{jheppub_kim}

\usepackage{pdflscape}
\usepackage{amsmath}
\usepackage{amssymb}
\usepackage{dcolumn}
\usepackage{bm}
\usepackage{color}
\usepackage{epsfig}
\usepackage{amsfonts}

\newcommand{\be}{\begin{equation}}
\newcommand{\ee}{\end{equation}}
\newcommand{\bea}{\begin{eqnarray}}
\newcommand{\eea}{\end{eqnarray}}

\def\be{\begin{equation}}
\def\ee{\end{equation}}
\def\bea{\begin{eqnarray}}
\def\eea{\end{eqnarray}}

\def \pd {\partial}

\begin{document}

\title{Cosmological behavior in extended nonlinear massive
gravity}

\author[a]{Genly Leon}
\author[a]{Joel Saavedra}
\author[b,a]{and Emmanuel N. Saridakis}

\affiliation[a]{Instituto de F\'{\i}sica, Pontificia Universidad  Cat\'olica de Valpara\'{\i}so, Casilla 4950, Valpara\'{\i}so, Chile}
\affiliation[b]{Physics Division, National Technical University of Athens, 15780 Zografou Campus,  Athens, Greece}

\emailAdd{genly.leon@ucv.cl}
\emailAdd{joel.saavedra@ucv.cl}
\emailAdd{Emmanuel$_-$Saridakis@baylor.edu}


\abstract{We perform a detailed dynamical analysis of various cosmological
scenarios in extended (varying-mass) nonlinear massive gravity. Due to the
enhanced freedom in choosing the involved free functions, this cosmological
paradigm allows for a huge variety of solutions that can attract the
universe at late times, comparing to scalar-field cosmology or usual
nonlinear massive gravity. Amongst others, it accepts quintessence,
phantom, or cosmological-constant-like late-time solutions, which moreover
can alleviate the coincidence problem. These features seem to be general
and non-sensitive to the imposed ansantzes and model parameters, and thus
extended nonlinear massive gravity can be a good candidate for the
description of nature. }

\maketitle

\newpage
\section{Introduction}

The idea of adding mass to the graviton is quite old \cite{Fierz:1939ix},
but the straightforward linear approach leads to the van Dam, Veltman,
Zakharov (vDVZ) discontinuity \cite{vanDam:1970vg,Zakharov:1970cc}, that is
the zero-mass limit of the obtained results does not provide the General
Relativity results. This is due to the fact that not all the extra degrees
of freedom, introduced by the graviton mass, decouple at the zero-mass
limit, since the longitudinal graviton preserves a finite coupling
to the trace of the energy-momentum tensor. This discontinuity can be
removed if one incorporates nonlinear terms \cite{Vainshtein:1972sx},
however it was soon realized that these necessary nonlinear terms introduce
the Boulware-Deser (BD) ghost degree-of-freedom  \cite{Boulware:1973my},
making the theory unstable. 

However, recently, a specific nonlinear extension of massive gravity
was formulated in \cite{deRham:2010ik,deRham:2010kj},   requiring
the 
Boulware-Deser ghost to be systematically removed (see
\cite{Hinterbichler:2011tt} for a review). Such a construction is
interesting at the theoretical level, since adding mass to a spin-two
particle is a well-defined problem by itself, however it has an additional
motivation, namely it is a new class of (Infra-Red) gravity modification
hoping to account for inflation and late-time acceleration. The theoretical
and phenomenological advantages led to a  significant amount of relevant
research
\cite{Koyama:2011yg,Hassan:2011hr,deRham:2011rn,CuadrosMelgar:2011yw,
Hassan:2011zd,Kluson:2011qe,Gumrukcuoglu:2011ew,
Volkov:2011an,vonStrauss:2011mq,Comelli:2011zm,Hassan:2011ea,
Berezhiani:2011mt,Gumrukcuoglu:2011zh,Khosravi:2011zi,Brihaye:2011aa,
Buchbinder:2012wb,Ahmedov:2012di,Bergshoeff:2012ud,Crisostomi:2012db,
Paulos:2012xe,Hassan:2012qv,Comelli:2012vz,Sbisa:2012zk,Kluson:2012wf,
Tasinato:2012mf,Morand:2012vx,Cardone:2012qq,Baccetti:2012bk,Gratia:2012wt,
Volkov:2012cf,deRham:2012kf,
Berg:2012kn,D'Amico:2012pi,Fasiello:2012rw,D'Amico:2012zv,
Baccetti:2012ge,Gong:2012yv,Volkov:2012zb,Nojiri:2012zu,
Deffayet:2012nr,
Chiang:2012vh,Hassan:2012wr,Kuhnel:2012gh,Motohashi:2012jd,Deffayet:2012zc,
Lambiase:2012fv,Gumrukcuoglu:2012wt,Gabadadze:2012tr,Kluson:2012zz,
Tasinato:2012ze,Gong:2012ny,Zhang:2012ap,Park:2012ds,Cai:2012db,
Wyman:2012iw,Burrage:2012ja,Nojiri:2012re,Park:2012cq,Alexandrov:2012yv,
deRham:2012ew,Hinterbichler:2013dv,Langlois:2012hk}.

Despite the successes of nonlinear massive gravity, it was realized that
the usual simple homogeneous and isotropic cosmological solutions are
unstable at the perturbation level \cite{DeFelice:2012mx}, which led to
less symmetric models \cite{D'Amico:2011jj,Gumrukcuoglu:2012aa}. However,
in \cite{Huang:2012pe} a different approach was followed, namely to
suitably extend the theory allowing for a varying graviton mass, driven by
a scalar field. This extended (varying-mass) nonlinear massive gravity
proves to exhibit interesting cosmological behavior, leading the universe
to lie at the quintessence or phantom regime, experience the phantom-divide
crossing \cite{Saridakis:2012jy}, or exhibit bouncing and cyclic behavior
\cite{Cai:2012ag}.

Since extended (varying-mass) nonlinear massive gravity exhibits
interesting phenomenological features when applied to cosmology, in the
present work we desire to perform a detailed dynamical analysis of such a
scenario. In this way we can bypass the complexities of the equations,
which prevent any complete analytical treatment, and investigate in a
systematic way the huge class of possible late-time cosmological behaviors,
calculating various observable quantities, such as the dark energy density
and equation-of-state parameters and the deceleration parameter.

The plan of the work is the following: In section \ref{model} we briefly
review the extended nonlinear massive gravity and its cosmological 
paradigm. In section \ref{Dynamicalanalysis} we perform a dynamical
analysis of both flat and open geometries, and in section
\ref{implications} we discuss the cosmological implications and the
physical behavior of the scenario. Finally, section \ref{Conclusions}
is devoted to the summary of the obtained results.

\section{Cosmology in extended nonlinear massive gravity}
\label{model}

In this section we briefly review cosmology in extended nonlinear massive
gravity \cite{Huang:2012pe, Saridakis:2012jy}. In this gravitational
framework the graviton mass is generalized to be varying, driven by a
scalar field.  The total action is written as
 \begin{eqnarray} \label{action0}
  S= \int d^4x \sqrt{-g} \left[ \frac{M_P^2}{2} R + V(\psi) ( U_2 +
\alpha_3 U_3 + \alpha_4 U_4)
-  \frac{1}{2} \pd_\mu \psi \pd^\mu
\psi -W(\psi)
\right]+S_m~,\ \ \
\end{eqnarray}
where $M_p$ is the reduced Planck mass, $R$ is the Ricci scalar,
$\psi$
is the extra canonical scalar field with $W(\psi)$ its usual potential
and $V(\psi)$ an additional potential coupling to the graviton potentials,
and
$\alpha_3$ and $\alpha_4$ are dimensionless parameters. 
The
graviton potentials write as
\begin{align}
U_2  =  \mathcal{K}^\mu_{[\mu}\mathcal{K}^\nu_{\nu]} ~, \quad
U_3  = \mathcal{K}^\mu_{[\mu}\mathcal{K}^\nu_{\nu}\mathcal{K}^\rho_{\rho]}
~, \quad
U_4  =
\mathcal{K}^\mu_{[\mu}\mathcal{K}^\nu_{\nu}\mathcal{K}^\rho_{\rho}\mathcal{
K}^\sigma_{\sigma]} ~,
\end{align}
with
\begin{eqnarray}
\mathcal{K}^\mu_\nu \equiv \delta^\mu_\nu-\sqrt{g^{\mu\rho}f_{AB}\pd_\rho
\phi^A
\pd_\nu \phi^B },
\end{eqnarray}
where we use the notation
\begin{eqnarray}
 \quad \mathcal{K}^\mu_{[\mu} \mathcal{K}^\nu_{\nu]}
\equiv \frac{1}{2}(\mathcal{K}^\mu_{\mu} \mathcal{K}^\nu_{\nu}
-\mathcal{K}^\mu_{\nu}\mathcal{K}^\nu_{\mu}) ~,
\end{eqnarray}
and similarly for the other antisymmetric expressions. Furthermore,
$f_{AB}$ is a fiducial metric, and     $\phi^A(x)$
are the St\"{u}ckelberg scalars introduced to restore general
covariance \cite{ArkaniHamed:2002sp}. The above extended scenario is still
free of the the BD ghost  
\cite{Huang:2012pe}. Finally, in order to obtain a realistic cosmology  in
  (\ref{action0}) we have allowed for the standard
matter action $S_m$, minimally-coupled to the dynamical metric,
corresponding to energy density $\rho_m$ and pressure $p_m$.

\subsection{Flat universe}
\label{flatmodel}

In order to extract the cosmological equations    we need to
consider specific ansatzes for the two metrics. For the physical metric we
assume a flat Friedmann-Robertson-Walker (FRW) form:
\begin{eqnarray}
d^2 s = -N(t)^2 d t^2 + a(t)^2 \delta_{ij} d x^i d x^j
~,
\end{eqnarray}
with $a(t)$ the scale factor and $N(t)$ the lapse function, and for
    the St\"{u}ckelberg fields we consider
\begin{eqnarray}
  \phi^0 = b(t)  ,  ~~~~\phi^i =a_{ref} x^i ~,
\end{eqnarray}
 with $a_{ref}$  
a (constant) reference scale factor.
We mention that contrary to standard massive gravity, where such a
choice for the
dynamical metric cannot be accompanied by a simple ansatz for the fiducial
one \cite{DeFelice:2012mx}, in the present extended scenario the
extra freedom does allow for a simple
Minkowski ansatz for the fiducial metric:
 \begin{eqnarray}
f_{AB}=\eta_{AB} .
\end{eqnarray}

Variation of the action with respect to $N$ and $a$ gives rise to
the   Friedmann equations
 \begin{eqnarray}
\label{Fr1}
3M_P^2 H^2& =& \rho_{DE}+\rho_m     ~,\\
\label{Fr2}
-2 M_P^2 \dot{H}& =&\rho_{DE}+p_{DE}+\rho_m +p_m   ~,
\end{eqnarray}
where we have defined the Hubble parameter $H=\dot{a}/a$, with
$\dot{a}=da/(Ndt)$, and finally we set $N=1$. In the
above
expressions we have defined the effective dark energy density and
pressure, incorporating the extra gravitational terms, as
\begin{eqnarray}
\label{rhomg22}
&&\rho_{DE} =\frac12 \dot{\psi}^2+W(\psi)+V(\psi)
\left(u-1\right)[f_3(u) +f_1(u)] \ \ \  \ \ \\
\label{pmg22}
&&p_{DE}  =\frac12 \dot{\psi}^2-W(\psi)-
V(\psi)f_4(u)-V(\psi)\dot{b}f_1(u) ~,
\end{eqnarray}
where
\begin{eqnarray}
&&f_1(u)
=3-2u+\alpha_3\left(3-u\right)\left(1-
u \right)+\alpha_4\left(1-u\right)^2
\nonumber \\
&&f_2(u) =
1-u+\alpha_3\left(1-u\right)^2+\frac{
\alpha_4}{3}
\left(1-u \right)^3
 \nonumber\\
&&f_3(u) =3-u+\alpha_3\left(1-u\right)
\nonumber\\
&&f_4(u) = -\left[6(1-u)+u^2+
\alpha_3\left(1-u\right)\left(4-2u
\right)
 +\alpha_4\left(1-u\right)^2\right] ~,
\label{fdefs00}
\end{eqnarray}
with 
\begin{eqnarray}
u=\frac{a_{ref}}{a}.
\end{eqnarray}
These satisfy the usual conservation equation
\begin{eqnarray}
 \dot{\rho}_{DE} +3H(\rho_{DE}+p_{DE})=0,
\end{eqnarray}
and moreover we can define the dark-energy equation-of-state parameter
as
\begin{eqnarray}
\label{wdedef}
w_{DE}\equiv \frac{p_{DE}}{\rho_{DE}}.
\end{eqnarray}

Variation of the  action  (\ref{action0}) with respect to the scalar field
$\psi$ provides
its evolution equation:
\begin{eqnarray}
\label{psievol}
 \ddot{\psi}+3H\dot{\psi}+\frac{d W}{d \psi}
+\frac{d V}{d\psi}
\left\{\left(u-1\right)
[f_3(u)+f_1(u)]+3\dot{b}
f_2(u)\right\}  =0 ~.\ \ \ \ \ \
\end{eqnarray}
Additionally, variation of (\ref{action0})  with respect to
$b$ provides the constraint equation
\begin{eqnarray}
\label{constraint}
 V(\psi)Hf_1(u)+\dot{V}(\psi)f_2(u) =0 ~.
\end{eqnarray}
 Finally, one must also consider the matter evolution
equation $\dot{\rho}_m +3H(\rho_m+p_m)=0$. In the following we assume  
matter to have a general equation-of-state parameter $w_m=\gamma-1\equiv
p_m/\rho_m$, where $\gamma$ is the barotropic index, focusing on the usual
dust case ($\gamma=1$) only when necessary.

The above cosmological application in a flat universe, although it leads to
interesting phenomenology, it has significant theoretical disadvantages.
These arise
mainly from the constraint equation (\ref{constraint}), which using 
(\ref{fdefs00}) in
general
gives \cite{Saridakis:2012jy}:
\begin{equation}
\label{constraint2aa}
 V(\psi(t)) = V_0\,e^{-\int \frac{f_1(u(a))}{a f_2(u(a))} d a}
 =
\frac{V_0
a_{ref}^3}{(a-a_{ref})[
\alpha_4a_{ref}^2-(3\alpha_3+2\alpha_4)aa_{ref}+(3+3\alpha_3+\alpha_4)a^2]
} ~.\ \ \ \ \ \
\end{equation}
As we observe this relation severely restricts the allowed
coupling-potential $V(\psi)$. Additionally, as we can see the varying
graviton square mass $V(\psi)$ diverges and changes sign at least for one
finite scale factor (namely at $a_{ref}$), independently of the model
parameters, and this would make the scenario unstable at the perturbation
level. Although one can still choose $a_{ref}$ at far past
($a_{ref}\lesssim10^{-9}$) in order to be smaller than the Big Bang
nucleosynthesis scale factor and not interfere with the standard
thermal history of the universe, or at the far future, or even  ``shield''
$a_{ref}$ with a cosmological bounce, case in which the universe is always
away from it \cite{Cai:2012ag}, such considerations can only cure the
problem phenomenologically, since at the theoretical level it remains
unsolved. Clearly, the scenario of a flat universe has a serious
disadvantage and therefore one should try to construct generalizations in
which these problems are absent. This will be performed in the next
subsection, where the addition of curvature makes the graviton mass square
always positive.

\subsection{Open universe}

Let us now consider an open\footnote{Similarly to usual massive
gravity, closed
FRW solutions are not possible since the fiducial Minkowski metric cannot
be foliated by closed slices  \cite{Gumrukcuoglu:2011ew,Huang:2012pe}.} FRW
form for the physical metric
\cite{Huang:2012pe}:
\begin{eqnarray}
d^2 s = -N(t)^2 d t^2 + a(t)^2 \delta_{ij} d x^i d x^j
- a(t)^2\frac{k^2(\delta_{ij}x^idx^j)^2}{1+k^2(\delta_{ij}x^ix^j)}~
,
\end{eqnarray}
with $N(t)$ the lapse function and $a(t)$ the scale factor, and $K<0$ with
$k=\sqrt{|K|}$. For the St\"{u}ckelberg fields we choose for simplicity
the forms  
\begin{eqnarray}
  \phi^0 = b(t)\sqrt{1+k^2(\delta_{ij}x^ix^j)}  ,  ~~~~\phi^i =k b(t) x^i
~.
\end{eqnarray}
Note that in this case there is no need for the introduction of a
reference scale factor $a_{ref}$, since it has been absorbed in $b(t)$.
Lastly, similarly to the flat case for the fiducial we consider
 \begin{eqnarray}
f_{AB}=\eta_{AB} .
\end{eqnarray}

Variation of the action (\ref{action0}) with respect to $N$ and $a$ gives
rise to
the following Friedmann equations
 \begin{eqnarray}
\label{Fr1b}
3M_P^2 \left(H^2-\frac{k^2  }{a^2}\right)& =& \rho_{DE}+\rho_m
~,\\
\label{Fr2b}
-2 M_P^2\left( \dot{H}+\frac{k^2  }{a^2}\right)&
=&\rho_{DE}+p_{DE}+\rho_m +p_m   ~.
\end{eqnarray}
In the
above
expressions we have defined the effective dark energy density and pressure
  as
\begin{eqnarray}
\label{rhomg22b}
&&\rho_{DE} =\frac12 \dot{\psi}^2+W(\psi)+V(\psi)
\left(X-1\right)[f_3(X) +f_1(X)] \ \ \  \ \ \\
\label{pmg22}
&&p_{DE}  =\frac12 \dot{\psi}^2-W(\psi)-
V(\psi)f_4(X)-V(\psi)\dot{b}f_1(X) ~,
\end{eqnarray}
but now the relevant functions become
\begin{eqnarray}
&&f_1(X)
=3-2X+\alpha_3\left(3-X\right)\left(1-
X \right)+\alpha_4\left(1-X\right)^2
\nonumber \\
&&f_2(X) =
1-X+\alpha_3\left(1-X\right)^2+
\frac{\alpha_4}{3}
\left(1-X \right)^3
 \nonumber\\
&&f_3(X) =3-X+\alpha_3\left(1-X\right)
\nonumber\\
&&f_4(X) = -\left[6-6X+X^2+
\alpha_3\left(1-X\right)\left(4-2 X
\right)
 +\alpha_4\left(1-X\right)^2\right] ~,
\label{fdefsopen}
\end{eqnarray}
where
\begin{eqnarray}
X=\frac{k b}{a}.
\end{eqnarray}
These verify the usual conservation equation
\begin{eqnarray}
 \dot{\rho}_{DE} +3H(\rho_{DE}+p_{DE})=0.
\end{eqnarray}

Variation of (\ref{action0}) with respect to the scalar field $\psi$
provides
its evolution equation:
\begin{eqnarray}
\label{psievolb}
 \ddot{\psi}+3H\dot{\psi}+\frac{d W}{d \psi}
+\frac{d V}{d\psi}
\left\{\left(X-1\right)
[f_3(X)+f_1(X)]+3\dot{b}
f_2(X)\right\}  =0 ~.\ \ \ \ \ \
\end{eqnarray}
Furthermore, variation  with respect to
$b$ provides the constraint equation
\begin{eqnarray}
\label{constraintb}
 V(\psi)\left(H-\frac{k  }{a}\right)f_1(X)+\dot{V}(\psi)f_2(X) =0 ~.
\end{eqnarray}
Finally, we consider also the matter conservation equation $\dot{\rho}_m
+3H(\rho_m+p_m)=0$.

\section{Dynamical analysis}
   \label{Dynamicalanalysis}

In order to investigate the   
cosmological behavior of the scenario of extended nonlinear massive
gravity we have to perform its dynamical analysis, and thus we
have to transform the involved cosmological
equations into the autonomous form $\label{eomscol}
\textbf{X}'=\textbf{f(X)}$
\cite{Copeland:1997et,Ferreira:1997au,Gong:2006sp,Chen:2008ft,Leon:2009rc},
where $\textbf{X}$ is the column vector of suitably introduced
auxiliary variables, $\textbf{f(X)}$ the corresponding  column
vector of the autonomous equations, and a prime denotes the derivative
with respect to $\ln a$. The critical points $\bf{X_c}$ are extracted
through $\bf{X}'=0$, and in order to examine their stability properties we
expand around $\bf{X_c}$ as $\bf{X}=\bf{X_c}+\bf{U}$, with $\textbf{U}$ the
corresponding perturbations of the variables. Thus, at the linear
perturbation level and for each critical point we find
$\label{perturbation} \textbf{U}'={\bf{Q}}\cdot
\textbf{U}$, where the matrix ${\bf {Q}}$ contains the coefficients of the
perturbation equations. Therefore, the eigenvalues of ${\bf {Q}}$ determine
the type and stability of the specific critical point.

The scenario at hand, that was presented in the previous section, consists
of the equations (\ref{Fr1}), (\ref{Fr2}) or (\ref{psievol}) and
(\ref{constraint2aa}) for the flat geometry, and (\ref{Fr1b}), (\ref{Fr2b})
or (\ref{psievolb}) and (\ref{constraintb}) for the open geometry, with
$\alpha_3$, $\alpha_4$ the model parameters. Although, as we discussed, the
flat case has theoretical disadvantages, for completeness in the following
we analyze it too, since it could still be cosmologically valid in
suitable frameworks, for example embedded into bouncing evolution.

As we can see, there are three unknown functions involved,
namely the usual
scalar potential $W(\psi)$, the varying graviton mass square $V(\psi)$ and
the St\"{u}ckelberg-field function $b(t)$. However, due to the
constraint equation, only two out of these three functions are free and can
be considered as input, while the third one is extracted from the
equations of motion. As
usual,  $W(\psi)$ is the one function that is always imposed by hand.
Throughout the work
we will consider the usual scalar field potential to have the
well-studied exponential form
\cite{Copeland:1997et,Ferreira:1997au,Gong:2006sp,Chen:2008ft}
\begin{equation}
W(\psi)=W_0 e^{-\lambda_W \psi}\label{Wpot}.
\end{equation}	
Thus, in the scenario at hand one could additionally either impose 
$V(\psi)$ at will
and
leave  $b(t)$ to be determined by the equations of motion in order to
obtain a consistent solution, or impose $b(t)$ as an input and leave
$V(\psi)$ to be determined by the equations. Definitely, the first
approach is theoretically more robust, corresponding
to the usual Lagrangian description where the potentials are imposed as
inputs in the theory, and it is the one that is followed in all the works
on the subject, that is the St\"{u}ckelberg fields are always extracted
by the equations
\cite{DeFelice:2012mx,Gumrukcuoglu:2012aa,D'Amico:2011jj,Langlois:2012hk,
Huang:2012pe}.
 Therefore, in the following subsection we will perform the phase-space
analysis imposing $V(\psi)$ as an input. However, for completeness, in a
separate
subsection we will also present the (theoretically less interesting) case
where $b(t)$ is considered as an input.

\subsection{Imposing $V(\psi)$ at will}

For the graviton mass square, and in order to be phenomenologically
consistent, without loss of generality we   assume an exponential form 
\begin{equation}
V(\psi)=V_0 e^{-\lambda_V\psi}\label{Vpot}.
\end{equation}	
In this case the graviton mass is small (at the order of the current
Hubble parameter in order to drive the current acceleration
\cite{Hinterbichler:2011tt}) at late times, as required by observations,
while it could play a significant role in the early universe.
Additionally, note that in the special case where $\lambda_V=0$, the
scenario at hand in the open case corresponds to the usual (constant-mass)
  nonlinear massive gravity.

\subsubsection{Flat universe}

In order to transform the cosmological system  (\ref{Fr1}),
(\ref{Fr2}) or (\ref{psievol}) and
(\ref{constraint2aa}) into its autonomous form, we introduce the
dimensionless variables 
\begin{equation}
u=\frac{a_{ref}}{a},\, Y=\frac{W(\psi)}{3 H^2}, \,Z=\frac{V(\psi)}{3
H^2}\label{auxiliary1}.
\end{equation}	
Taking the derivatives of (\ref{auxiliary1}) and
using (\ref{Fr1}), (\ref{Fr2}) and (\ref{constraint2aa}), we obtain 
the evolution equations for $u, Y,$ and $Z$, that is the autonomous form
of the cosmological system, as
\begin{eqnarray}
&&u'=-u \nonumber\\
&&Y'=Y\left[2(1+q)-\frac{\lambda_W}{\lambda_V}\frac{f_1(u) }{f_2(u)
}\right]\nonumber\\
&&Z'=Z\left[2(1+q)-\frac{f_1(u) }{f_2(u)}\right]\label{ODEs1},
\end{eqnarray} where primes denote derivative  with respect to $\ln a$. In
the above expressions $q=-1-\frac{\dot H}{H^2}$ is the
deceleration parameter, and the involved $\dot{H}$ can be expressed in
terms of the auxiliary variables as
\begin{align}
\label{eq3.10}
&\dot H= \frac{H^2 g_1(u,Y,Z)}{6 \lambda_V^2 f_2^3-f_1^2 f_2} , 
\end{align}
with
\begin{eqnarray}
&&g_1(u,Y,Z)=u f_1 \left(f_1 \frac{d f_2}{d u}-f_2 \frac{d f_1}{d
u}\right) +3
   \lambda_V^2 Z f_2^2 \left\{3 f_2 f_4-(u-1)
   (f_1+f_3) [f_1-3 (\gamma -1) f_2]\right\} \nonumber \\ 
&&\ \ \ \ \ \ \ \ \  \ \   \ \ \ \ \ \  +3  f_1^2 f_2+3
\lambda_V Y f_2^2 (3
   \gamma  \lambda_V f_2-\lambda_W f_1)-9 \gamma 
   \lambda_V^2 f_2^3+  (\gamma -2)  \frac{3}{ 2
   }f_1^2 f_2 ,
\end{eqnarray}
as it arises from  \eqref{Fr2} and \eqref{psievol} through
elimination of  $\dot{b}$ (for simplicity we have omitted the argument
$u$ in $f_1(u)$ and
$f_2(u)$). On the other hand, $\dot{H}$ elimination
between \eqref{Fr2} and \eqref{psievol} gives
\begin{align}\label{eq3.10bbb}
	3 Z \dot b= \frac{g_2(u, Y,Z)}{6 \lambda_V^2 f_2^3-f_1^2 f_2} ,
\end{align}
with 
\begin{eqnarray}
&&g_2(u,Y,Z)=	-2 u f_2 \frac{d f_1}{du}+f_1 \left(2 u \frac{d f_2}{du}-3
\gamma 
   f_2\right) +6   f_1  f_2   +3 Y f_2 (\gamma 
   f_1-2 \lambda_V \lambda_W f_2) \ \ \ \  \ \  \ \  \nonumber \\ 
&&\ \    \ \ \ \  \ \   \ \ \ \  \ \ \ \ \   +3 Z f_2 \left\{(u-1)
(f_1+f_3)
   \left[(\gamma -1) f_1-2 \lambda_V^2 f_2\right]+f_1
   f_4\right\}  + \frac{  (\gamma -2)}{2
\lambda_V^2 f_2} f_1^3  .
\end{eqnarray}

 Furthermore, using  \eqref{Fr1} we can
express the dark energy density parameter
$\Omega_{DE}\equiv\frac{\rho_{DE}}{3 H^2}$  in terms of the auxiliary
variables  as 
\begin{equation}\label{OmegaDE1}
\Omega_{DE} =\frac{f_1^2}{6 \lambda_V^2 f_2^2}+(u-1) Z
   \left[f_1(u)+f_3(u)\right]+Y,
\end{equation}
while using (\ref{eq3.10}) and
(\ref{eq3.10bbb}) we can express the  dark
energy equation-of-state parameter  and the  deceleration parameter  
respectively as
\begin{equation}
w_{DE}=\frac{-Z \left\{\frac{f_1(u) g_2(u, Y,Z)}{3 Z\left[6 \lambda_V^2
f_2^3-f_1^2
f_2\right]}+f_4(u)\right\}+\frac{f_1(u)^2}{6
\lambda_V^2 f_2(u)^2}-Y}{(X-1) Z [f_1(u)+f_3(u)]+\frac{f_1(u)^2}{6
\lambda_V^2 f_2(u)^2}+Y},
\label{wdecase1}
\end{equation}
\begin{align}
\label{qcase1}
q=-1-\frac{g_1(u,Y,Z)}{6 \lambda_V^2 f_2^3-f_1^2 f_2}.
\end{align}

In summary,  \eqref{ODEs1} accounts for an autonomous system defined in the
phase space
\begin{equation}
 \left\{(u,Y,Z): 0\leq \frac{f_1(u)^2}{6 \lambda_V^2 f_2(u)^2}+(u-1) Z
   [f_1(u)+f_3(u)]+Y\leq 1,       u\geq 0, Y\geq 0,
Z\geq 0 \right\},
\end{equation}
as it arises from the physicality requirements $a\geq0$, $V(\psi)\geq0$,
$W(\psi)\geq0$ and $0\leq\Omega_{DE}\leq1$, and it
  is in general non-compact.

The real and physically meaningful critical points $(u_c,Y_c,Z_c)$ of the
autonomous system  \eqref{ODEs1} (that is corresponding to  
$0\leq\Omega_{DE}\leq1$), are
obtained by setting the left-hand-sides of these equations  to zero, and
they are presented in Table \ref{Tab1}, along with their existence
conditions. For each critical point we calculate the $3\times3$
matrix ${\bf {Q}}$ of the linearized perturbation equations of the system
\eqref{ODEs1}, and  examining the sign of the real part of the eigenvalues
of ${\bf {Q}}$ we determine the type and stability
of this point. The details of the analysis and the various eigenvalues
are presented in Appendix \ref{App1}, and in Table \ref{Tab1} we
summarize the stability results (note that in the case of standard matter
($\gamma=1$)     points 
$P_1$ and $P_2$ belong to a curve of critical points). Finally, using
(\ref{OmegaDE1}), (\ref{wdecase1}) and (\ref{qcase1}), for each critical
point we calculate the corresponding values of $\Omega_{DE}$,
$w_{DE}$, and $q$.
\begin{table*}[!]
\begin{center}
\resizebox{\columnwidth}{!}
{\begin{tabular}{|c|c|c|c|c|c|c|c|c|}
\hline
&&&& &&&&  \\
{\small{ Cr. P.}}& $u_c$ & $Y_c$ & $Z_c$ & Existence&  Stable for &
$\Omega_{DE}$& $w_{DE}$ &q \\
\hline \hline
$P_1$& 0 & 0 & 0  & for $\lambda_V^2\geq
\frac{3}{2}$&  {\small{  
$\gamma<\min\left\{1,\frac{\lambda_W}{\lambda_V}\right\},\lambda_V^2\geq
\frac{3}{2}$  }}  & $\frac{3}{2 \lambda_V^2}$ & $\gamma -1$ &$\frac{3
\gamma }{2}-1$
\\[0.2cm]
&&&& & saddle point otherwise  &&   &\\ \hline
$P_2$& 0 & 0 &  $\frac{3-2 \lambda_V^2}{2\mu \lambda_V^2 }$ 
&{\small{$\mu> 0, 0<\lambda_V^2\leq
\frac{3}{2}$ }}   &  
$\gamma>1,\frac{\lambda_W}{\lambda_V}>1$  & $1$ & 0 & $\frac{1}{2}$ 
\\[0.2cm]
&&&&    or $\mu< 0, \lambda_V^2\geq
\frac{3}{2}$ & saddle point otherwise &&&   \\[0.2cm] \hline
$P_3$ & 0 &  {\small{ $1-\frac{3}{2 \lambda_V^2}$}} & 0 & for
$\lambda_V^2\geq
\frac{3}{2}$  &{\small{ $\frac{\lambda_W}{\lambda_V}<\min
\left\{1,\gamma\right\}, \lambda_V^2\geq \frac{3}{2}$}}  & 1 &
{\small{$\frac{\lambda_W}{\lambda_V}-1$}} & {\small{$\frac{3 \lambda_W}{2
   \lambda_V}-1$}}
\\[0.2cm]
&&&& &  saddle point otherwise &&  & \\ 
 \hline
\end{tabular}}
\end{center}
\caption[crit]{\label{Tab1} The real and physically meaningful critical
points of the autonomous system  (\ref{ODEs1}), their existence and
stability conditions, and the corresponding values of the dark-energy
density parameter $\Omega_{DE}$, of the  dark-energy equation-of-state
parameter $w_{DE}$, and of the deceleration parameter $q$.  We have
introduced the notation $\mu=(4 \alpha_3+\alpha_4+6)$. }
\end{table*}

\subsubsection{Open universe}
\label{Vgivenopen}

In order to transform the cosmological system (\ref{Fr1b}), (\ref{Fr2b})
or (\ref{psievolb}) and (\ref{constraintb})  into its autonomous form, we
introduce the
dimensionless variables 
\begin{equation}
X=\frac{k   b}{a},\, Y=\frac{W(\psi)}{3 H^2},\,Z=\frac{V(\psi)}{3
H^2},\,U=\frac{\dot\psi}{\sqrt{6} H},\,\Omega_k=\frac{  k}{a H}.
\label{auxiliary2}
\end{equation}	
Differentiating with respect to $\ln a$ we obtain  the autonomous
form of the cosmological system:
\begin{align}\label{ODEs2}
&X'=-X+\Omega_k \dot b\nonumber \\
& Y'=Y \left[2 (q+1)-\sqrt{6} \lambda_W U\right]\nonumber \\
&Z'=Z \left[2 (q+1)-\sqrt{6} \lambda_V U\right]\nonumber \\
& U'=3 \sqrt{\frac{3}{2}} \lambda_V
   Z f_2 \dot b+\frac{1}{2} \left\{\sqrt{6} \left[\lambda_V (X-1) Z
(f_1+f_3)+\lambda_W Y\right]+2 (q-2) U\right\}\nonumber \\
&\Omega_k'=q \Omega_k,
\end{align}
with $q=-1-\frac{\dot H}{H^2}$, and 
where  for simplicity we have omitted the argument $X$ in $f_1(X)$ and
$f_2(X)$. In the above expressions  $\dot{H}$
and $\dot{b}$ are given by \eqref{Fr2b} and \eqref{psievolb} as
\begin{align}\label{eq42}
&\dot H=\frac{H^2 g_1(X,Y,Z,U,\Omega_k,H^2)}{\lambda_V \left\{-2
(\Omega_k-1)
\Omega_k f_2 \frac{d f_1}{d X}+2 (\Omega_k-1) \Omega_k f_1
   \frac{d f_2}{d X}+3 Z f_1^2 f_2 \left[6 (\Omega_k-1)^2
H^2+1\right]\right\}}   \\
\label{eq42bb}
&	\dot b=\frac{ g_2(X,Y,Z,U,\Omega_k,H^2)}{\lambda_V \left\{-2
(\Omega_k-1) \Omega_k f_2 \frac{d f_1}{d X}+2 (\Omega_k-1) \Omega_k f_1
   \frac{d f_2}{d X}+3 Z f_1^2 f_2 \left[6 (\Omega_k-1)^2
H^2+1\right]\right\}},
\end{align}
with
\begin{eqnarray}\label{g1}
&& g_1(X,Y,Z,U,\Omega_k,H^2)=\frac{d f_1}{d X} \left\{-3 \lambda_V
(\Omega_k-1) Z f_1 f_2 \left[(\gamma -1) (X-1)
\Omega_k+X\right] \right.\nonumber \\
&& \ \  \left.  -\lambda_V (\Omega_k-1)
   \Omega_k f_2 \left\{3 Z \left[(\gamma -1) (X-1) f_3+f_4\right]+3 (\gamma
-2) U^2+3
 \gamma  \left(Y+\Omega_k^2-1\right)-2 \Omega_k^2\right\}\right\}
\nonumber
\\
&& \ \  +\frac{d f_2}{d X} \left\{3 \lambda_V (\Omega_k-1) Z f_1^2
\left[(\gamma -1) (X-1) \Omega_k+X\right]
\right.\nonumber \\  
&& \ \ 
\left.-\lambda_V
   (\Omega_k-1) \Omega_k f_1 \left[-3 (\gamma -1) (X-1) Z f_3-3 Z f_4-3
(\gamma -2) U^2-3 \gamma   \left(Y+\Omega_k^2-1\right)+2
\Omega_k^2\right]\right\}
\nonumber\\
&& \ \ +f_1^2 f_2 
\left\{9 \lambda_V
(\Omega_k-1)^2 Z H^2 \left\{3 Z \left[(\gamma -1) (X-1)
   f_3+f_4\right]+3 (\gamma -2) U^2  \right. \right.
 \nonumber
 \\ 
&& \ \ \left.
\left. +3 \gamma  \left(Y+\Omega_k^2-1\right)-2 \Omega_k^2\right\}-3
\lambda_V \Omega_k
   Z\right\}
\nonumber\\ 
&& \ \  +f_1^3 \left\{27 (\gamma -1) \lambda_V (X-1)
(\Omega_k-1)^2 Z^2 f_2 H^2  \right.
 \nonumber \\ 
&& \ \ \left.+9 (\Omega_k-1)^2
Z H^2
   \left[-\lambda_V (X-1) Z f_3+\sqrt{6} U-\lambda_W Y\right]\right\}
\nonumber\\ 
&& \ \  -9 \lambda_V (X-1) (\Omega_k-1)^2 Z^2 f_1^4 H^2
\end{eqnarray}
\begin{eqnarray}
&&g_2(X,Y,Z,U,\Omega_k,H^2)= -2 \lambda_V X (\Omega_k-1) f_2 \frac{d f_1}{d
X}+2 \lambda_V X (\Omega_k-1) f_1 \frac{d f_2}{d X} 
\nonumber \\
&& \ \  -\lambda_V
   f_1 f_2 \left\{3 Z \left[(\gamma -1) (X-1) f_3+f_4\right]+3 (\gamma -2)
U^2+3 \gamma \left(Y+\Omega_k^2-1\right)-2 (\Omega_k-1)
   \Omega_k\right\} 
\nonumber \\ 
&& \ \  +f_1^2 \left\{6 (\Omega_k-1)^2 H^2
\left[-\lambda_V (X-1) Z f_3+\sqrt{6} U-\lambda_W Y\right]-3
   (\gamma -1) \lambda_V (X-1) Z f_2\right\}
\nonumber \\ 
&& \ \  -6 \lambda_V
(X-1) (\Omega_k-1)^2 Z f_1^3 H^2,	
\end{eqnarray}
where  $H^2$ is given from \eqref{constraintb} as 
\begin{equation}\label{eq41}
H^2=\left[\frac{\lambda_V f_2(X)}{\sqrt{6} (1-\Omega_k) f_1(X) U}\right]^2.
\end{equation}

\begin{table*}[!]
\begin{center}
\resizebox{\columnwidth}{!}{
\begin{tabular}{|c|c|c|c|c|c|c|}
\hline
&&&& && \\
 Cr. P.& $X_c$ & $Y_c$ & $Z_c$ & $U_c$ & $\Omega_{kc}$ & Exists for  \\
\hline \hline
$Q_1$& 0&  $\frac{\lambda_V}{\lambda_V-\lambda_W}$ & $Z_c$ &  0 & 0   &
$0\leq \frac{\lambda_V}{\lambda_V-\lambda_W}-\mu Z_c\leq 1$
\\[0.2cm]\hline
$Q_2$& 0& 0& $Z_c$ & $\frac{\sqrt{\frac{3}{2}}}{\lambda_V}$ & 0   & $0\leq
\frac{3}{2 \lambda_V^2}-\mu Z_c\leq 1$ 
\\[0.2cm]\hline
$Q_3$& 0 & 0 & $Z_c$ & $\frac{\lambda_V}{\sqrt{6}}$ & 0 & $0\leq
\frac{\lambda_V^2}{6}-\mu Z_c\leq 1$ 
\\[0.2cm]\hline
$Q_4$& 0& 0&0 & $U_c$ &0 & $0\leq U_c^2\leq 1$  
\\[0.2cm]\hline
$Q_5$& 0 & $Y_{c 5}$ & 0& $U_c$ & 0 & $0\leq\frac{3 \lambda_V+6 \lambda_V
U_c^2-\sqrt{6} U_c
(\lambda_V \lambda_W+3)}{3 (\lambda_V-\lambda_W)}\leq 1$ 
\\[0.2cm]\hline
$Q_6^\pm$ &$X_c$ &  $\frac{4}{3\lambda_W^2}$ &0
&$\frac{\sqrt{6}}{3\lambda_W}$ &
$\pm\sqrt{1-\frac{2}{\lambda_W^2}}$  & $\lambda_W^2\geq 2$ 
\\[0.2cm]\hline
$Q_7^\pm$ &0&0 & $-\frac{4}{3 \lambda_V^2 \mu}$&
$\frac{\sqrt{6}}{3\lambda_V}$
& $\pm\sqrt{1-\frac{2}{\lambda_V^2}}$  & $\lambda_V^2\geq 2$ 
 \\[0.2cm]\hline
$Q_{8}$ &$1$&0 & 0& 0 & 1 & always 
\\[0.2cm]\hline
$Q_{9}$ &$\frac{2 \alpha_3+\alpha_4- \sqrt{4 \alpha_3^2-6
\alpha_4}}{\alpha_4}.$&0 & 0& 0 & 1 & $\alpha_3^2\geq \frac{3}{2}\alpha_4$
 
\\[0.2cm]\hline
$Q_{10}$ &$\frac{2 \alpha_3+\alpha_4+ \sqrt{4 \alpha_3^2-6
\alpha_4}}{\alpha_4}$&0 & 0& 0 & 1 & $\alpha_3^2\geq \frac{3}{2}\alpha_4$  
\\[0.2cm]\hline
$Q_{11}$ &$X_{c 11}$&0 & 0& 0 & $-1$ & $X_{c 11}\in\mathbb{R}$  
\\[0.2cm]\hline
\end{tabular}
}
\end{center}
\caption[crit]{\label{crit01open1} The real and physically meaningful
curves of critical points, and individual critical points, of the
autonomous system (\ref{ODEs2}) and their existence conditions, for the
case of dust matter ($\gamma=1$). We have
introduced the notations   
$\mu=(4
\alpha_3+\alpha_4+6),$ $Y_{c 5}=\frac{\sqrt{6} \lambda_V+U_c \left[-2
\lambda_V \lambda_W+\sqrt{6}U_c (\lambda_V+\lambda_W)-6\right]}{\sqrt{6}
(\lambda_V-\lambda_W)}$,  and $X=X_{c 11}$ is the unique real solution of the
equation  $-2 \alpha_3 \left(X^2+X-2\right)+\alpha_4 (X+1) (X-1)^2+6=0.$
}
\end{table*}
\begin{table*}[!]
\begin{center}
\resizebox{\columnwidth}{!}{
\begin{tabular}{|c|c|c|c|c|}
\hline
&&&&   \\
 Cr. P.  & Stability&
$\Omega_{DE}$& $w_{DE}$ &q \\
\hline \hline
$Q_1$  &
Non-Hyperbolic, 3D stable manifold  &
$\frac{\lambda_V}{\lambda_V-\lambda_W}-Z_c\mu$  & ${w_{DE}}_1$ & -1
\\[0.2cm]\hline
$Q_2$  & saddle point & $\frac{3}{2 \lambda_V^2}-\mu Z_c$ 
& $0$ & $\frac{1}{2}$
\\[0.2cm]\hline
$Q_3$  & Non-Hyperbolic, 3D stable manifold for  & $\frac{3}{2
\lambda_V^2}-\mu Z_c$ 
& $0$ & $\frac{1}{2}$
\\[0.2cm]
& $\lambda_V^2<2, \lambda_V(\lambda_V-\lambda_W)<0$  & & &
\\[0.2cm]
& saddle otherwise   & & &
\\[0.2cm]\hline
$Q_4$  &Non-Hyperbolic, 3D stable manifold for  & $U_c^2$  &
$\frac{3-\sqrt{6} \lambda_V U_c}{3 U_c^2-\sqrt{6} \lambda_V U_c}$ &
$\frac{-\sqrt{6} \lambda_V-3 \sqrt{6} \lambda_V U_c^2+12 U_c}{6 U_c-2
   \sqrt{6} \lambda_V}$
\\[0.2cm]
& $-1<U_c\leq -\frac{1}{\sqrt{3}}, \frac{2 \sqrt{6} U_c}{3
U_c^2+1}<\lambda_V<\sqrt{\frac{3}{2}} U_c,
\lambda_W<\lambda_W^*(U_c,\lambda_V)$ or  & & &
\\[0.2cm]
& $-\frac{1}{\sqrt{3}}<U_c<0, \sqrt{6} U_c<\lambda_V<\sqrt{\frac{3}{2}}
U_c, \lambda_W<\lambda_W^*(U_c,\lambda_V)$ or  & & &
\\[0.2cm]
& $0<U_c\leq \frac{1}{\sqrt{3}},   \sqrt{\frac{3}{2}}
U_c<\lambda_V<\sqrt{6} U_c, \lambda_W>\lambda_W^*(U_c,\lambda_V)$ or  & & &
\\[0.2cm]
& $\frac{1}{\sqrt{3}}<U_c<1,\sqrt{\frac{3}{2}} U_c<\lambda_V<\frac{2
   \sqrt{6} U_c}{3 U_c^2+1}, \lambda_W>\lambda_W^*(U_c,\lambda_V)$  & & &
\\[0.2cm]
& saddle otherwise   & & &
\\[0.2cm]\hline
$Q_5$  & Non-Hyperbolic, 2D stable manifold for  &
${\Omega_{DE}}_5$  & ${w_{DE}}_5$ & $\sqrt{\frac{3}{2}}
   \lambda_W U_c-1$ 
\\[0.2cm]
& $0\leq {\Omega_{DE}}_5\leq 1$ and $U_c \lambda_W
<\min\left\{\sqrt{\frac{3}{2}}, U_c\lambda_V\right\} $ & & &
\\[0.2cm]
& saddle otherwise   & & &
\\[0.2cm]\hline
$Q_6^\pm$  & Non-Hyperbolic, 4D stable manifold for &
$\frac{2}{\lambda_W^2}$ & $-\frac{1}{3}$ &0\\[0.2cm]
& $ 2<\lambda_W^2<\frac{8}{3}, \frac{\lambda_V}{\lambda_W}>1$ or & & &
\\[0.2cm]
& $\lambda_W^2>\frac{8}{3}, \frac{\lambda_V}{\lambda_W}>1$  & & &
\\[0.2cm]
& saddle otherwise   & & &
\\[0.2cm]\hline
$Q_7^\pm$  & Non-Hyperbolic, 4D stable manifold for &
$\frac{2}{\lambda_V^2}$ & $-\frac{1}{3}$ &0 \\[0.2cm]
&  $\begin{array}{c}\frac{\lambda_V}{\lambda_W}>1,
\lambda_V^2\geq 2, \\ \Re\left[\Delta_1(\alpha_3,\alpha_4,
\lambda_V,\lambda_W)\right]<0, \\\Re\left[\Delta_2(\alpha_3,\alpha_4,
\lambda_V,\lambda_W)\right]<0\end{array}$  & & &
\\[0.2cm]
& saddle otherwise   & & &
\\[0.2cm]\hline
$Q_{8}$& saddle point &0& $-\frac{1}{3}$ & 0 \\[0.2cm]\hline
$Q_{9}$  & saddle point & 0 &
$-1$ & 0  \\[0.2cm]\hline
$Q_{10}$  & saddle point & 0 &
$-1$ & 0 \\[0.2cm]\hline
$Q_{11}$ &  saddle point & 0 &
$w(X_{c 11})$ & 0 \\[0.2cm]\hline
\end{tabular}
}
\end{center}
\caption[crit]{\label{crit01open2} 
The stability conditions and the values
of the observables $\Omega_{DE}$,  $w_{DE}$ and $q$, for the real and
physically meaningful
curves of critical points, and individual critical points, of the
autonomous system (\ref{ODEs2}),   for the case of dust matter
($\gamma=1$). The notations are the same with Table \ref{crit01open1}.
Additionally, we have defined
${w_{DE}}_1=\frac{\lambda_V-\lambda_W}{\lambda_V
\left[Z_c (4 \alpha_3+\alpha_4+6)-1\right]-\lambda_W Z_c (4
\alpha_3+\alpha_4+6)},$ 
${w_{DE}}_5=\frac{3
   (\lambda_V-\lambda_W) \left[\sqrt{6} \lambda_V+\sqrt{6} \lambda_W
U_c^2-U_c (2 \lambda_V \lambda_W+3)\right]}{\left(3
   U-\sqrt{6} \lambda_V\right) \left[3 \lambda_V+6 \lambda_V U_c^2-\sqrt{6}
U_c (\lambda_V \lambda_W+3)\right]},$ ${\Omega_{DE}}_5=\frac{3 \lambda_V+6
\lambda_V U_c^2-\sqrt{6} U_c
(\lambda_V \lambda_W+3)}{3 (\lambda_V-\lambda_W)},$ 
$\lambda_W^*(U_c,\lambda_V)=\frac{\sqrt{6} \lambda_V^2
\left(U_c^2+1\right)-3 \lambda_V \left(U_c^2+3\right) U_c+3 \sqrt{6}
U_c^2}{U_c \left(2 \lambda_V^2+3 U_c^2-2 \sqrt{6} \lambda_V U_c\right)}$ and $w(X)=\frac{X \left[\alpha_3^2 (X-1)^2-2 \alpha_3
(X-1)-\alpha_4 (X-1)^2+3\right]}{\left[3 \alpha_3
(X-1)-\alpha_4 (X-1)^2-3\right] \left\{4
   \alpha_3+\alpha_4+X [\alpha_3 (X-5)+\alpha_4 (X-2)-3]+6\right\}}.$ The
symbol $\Re[z]$ denotes the real part of the complex number $z$.}
\end{table*}
\begin{table*}[!]
\begin{center}
\resizebox{\columnwidth}{!}
{\begin{tabular}{|c|c|c|c|c|c|c|c|c|c|c|}
\hline
&&&& &&&& &&  \\
{\small{ Cr. P.}}& $X_c$  & $Y_c$ & $Z_c$ & $U_c$  &  $\Omega_{k c}$ &
Exists for&  Stability &
$\Omega_{DE}$& $w_{DE}$ &q \\
\hline \hline
$Q_{12}$& 0 & 0 & 0  & $-1$ & 0    & always  & unstable & 1 & 1 & 2
\\[0.2cm]
\hline
$Q_{13}$& 0 & 0 & 0  & $1$ & 0    & always  & unstable & 1  & 1 & 1
\\[0.2cm]
\hline
$Q_{14}$& 0 & $1-\frac{\lambda_W^2}{6 }$ & 0  &
$\frac{\lambda_W}{\sqrt{6}}$ & 0    & $\lambda_W^2\leq 6$ & stable
node for $-\sqrt{2}<\lambda_W<0,\lambda_V<\lambda_W$ or   & 1 &
$-1+\frac{\lambda_W^2}{3}$ &  
$-1+\frac{\lambda_W^2}{2}$\\
&&&&&&&   $0<\lambda_W<\sqrt{2}, \lambda_V>\lambda_W$  &&&
\\[0.2cm]
&&&&&&&   saddle point otherwise  &&&
\\[0.2cm]
\hline
$Q_{15}$& 0 & $\frac{3}{2 \lambda_W^2}$ & 0  & $\sqrt{\frac{3}{2
\lambda_W^2}}$ & 0    &  $\lambda_W^2\geq 3$  & saddle point & $\frac{3}{2
\lambda_W^2}$ &  0 & $\frac{1}{2}$
\\[0.2cm]
\hline
$Q_{16}$& 0 & $0$ & 0  & $0$ & 0    & always  & saddle point & $0$ &  0 &
$\frac{1}{2}$
\\[0.2cm]
\hline
$Q_{17}$& 0 & $0$ & 0  & $0$ & $-1$    & always  & saddle point & $0$
& 0
& 0
\\[0.2cm]
\hline
$Q_{18}$& 0 & $0$ & 0  & $0$ & $1$    & always  & saddle point & $0$
& 0
& 0
\\[0.2cm]
\hline
\end{tabular}
} 
\end{center}
\caption[crit]{\label{quintessence} 
The interesting individual critical points of the curve of critical
points $Q_5$ of Table \ref{crit01open2},  their existence and
stability conditions, and the corresponding values of the observables
$\Omega_{DE}$,   $w_{DE}$, and $q$.}
\end{table*}

Furthermore, using  \eqref{Fr1b} we can
express the dark energy density parameter   in terms of the auxiliary
variables  as
\begin{equation}\label{OmegaDE2}
\Omega_{DE}\equiv\frac{\rho_{DE}}{3 H^2}=(X-1) Z
\left[f_1(X)+f_3(X)\right]+U^2+Y,
\end{equation}
while using (\ref{eq42bb}) we can express the  dark
energy equation-of-state parameter as 
\begin{equation}\label{DEEoS2}
w_{DE}=\frac{-Z \left[f_1(X) \dot b+f_4(X)\right]+U^2-Y}{(X-1) Z
[f_1(X)+f_3(X)]+U^2+Y},
\end{equation}
and finally  
 using
(\ref{eq42}),(\ref{eq41}) the  deceleration parameter  
is expressed as
\begin{align}
\label{qcase2}
& q= -1-\frac{g_1(X,Y,Z,U,\Omega_k,H^2)}{\lambda_V \left\{-2 (\Omega_k-1)
\Omega_k
f_2 \frac{d f_1}{d X}+2 (\Omega_k-1) \Omega_k f_1
   \frac{d f_2}{d X}+3 Z f_1^2 f_2 \left[6 (\Omega_k-1)^2
H^2+1\right]\right\}}.
\end{align}

In summary,  \eqref{ODEs2} accounts for an autonomous system, which its
physical part of   the
phase space ($a\geq0$, $V(\psi)\geq0$,
$W(\psi)\geq0$ and $0\leq\Omega_{DE}+\Omega_k^2\leq1$) is defined as
\begin{equation}
  \left\{(X,Y,Z,U,\Omega_k): 0\leq (X-1) Z \left[f_1(X)+f_3(X)\right]+U^2+Y
+\Omega_k^2\leq 1,   X\geq
0, Y\geq 0, Z\geq 0 \right\},
\end{equation}
which is in general non-compact. 

Let us extract the critical points of the autonomous system \eqref{ODEs2},
setting the left-hand-sides of these equations to zero. From the last
equation of \eqref{ODEs2} it follows that either $q=0$ or $\Omega_k=0$, and
therefore we  can simplify the investigation and examine these two cases
separately. The details of the analysis, the critical points and critical
curves, the various eigenvalues and the stability conditions are presented
in Appendix \ref{App2}, and in the Table \ref{crit01open1} we display the
real and physically meaningful critical points and their existence
conditions  for the most
interesting case of dust matter ($\gamma=1$), while in Table
\ref{crit01open2} we present their stability conditions and the values of
the observables $\Omega_{DE}$,
$w_{DE}$, and $q$ using (\ref{OmegaDE2}), (\ref{DEEoS2}) and
(\ref{qcase2}). 

We mention here, that the variable choice
(\ref{auxiliary2}) allows for an easy, partial, classification of
expanding and contracting solutions. In particular, solutions with  
$\Omega_k=k/(aH)>0$ correspond to $H>0$ and thus to expansion, while those
with $\Omega_k<0$ correspond to $H<0$ and therefore to contraction
($k=\sqrt{|K|}$ throughout this work). That is why points with
$\Omega_k>0$ are denoted with the subscript ``+'', while those with
$\Omega_k<0$ are denoted with the subscript ``-''. However, this is only 
a partial classification, since it cannot work  for solutions with
$\Omega_k=0$, which can be either expanding or contracting.
Furthermore, note that although our model admits expanding and 
contracting solutions, from the fifth equation of \eqref{ODEs2} we deduce
that the sign of $\Omega_k$ is invariant, and thus    
transitions from contracting to expanding solutions or vice
versa do not exist. Nevertheless, since such transitions do exist in the
flat geometry \cite{Cai:2012ag}, there could still exist in the non-flat
scenario at hand too, but at the edge of the phase space, which
could be revealed only through application of Poincar\'e
central projection method \cite{PoincareProj,Leon2011,Leon:2012mt}. This
analysis lies beyond the scope of the present work and it is left for
future investigation.

Finally, we stress that the curve of critical points $Q_5$ contains many
interesting individual points, and for that reason we display them
separately in Table \ref{quintessence}, along with their existence and
stability conditions and the corresponding values of the observables. Note
that these points contain the standard quintessence points 
\cite{Copeland:1997et,Copeland:2009be}, however the stability conditions
are slightly different, due to the presence of extra phase-space
dimensions, namely     curvature and   graviton mass.

\subsection{Imposing $b(t)$ at will}

In the previous subsection we performed the dynamical analysis
following the 
theoretically robust approach in Lagrangian descriptions, that is
imposing the potential $V(\psi)$ (graviton varying square mass) as an input
and letting the St\"{u}ckelberg field function $b(t)$ to
be determined by the equations. However, for completeness, and in order to
compare with similar studies in the literature \cite{Wu:2013ii}, in this
section we follow the theoretically less justified, alternative approach, that is to impose $b(t)$ at
will and let $V(\psi)$ be determined by the equations. Similarly to the
previous subsection, we will consider the flat and open geometry
separately, using different $b(t)$ ansantzes in the two case for
convenience.

\subsubsection{Flat universe}
 
In this case we impose   $b(t)= B t$  with $B>0$, since this leads  
to $\dot{b}=B$  which
simplifies significantly the cosmological equations. In the following  we
focus on the dust matter case ($\gamma=1$), however the analysis can be
straightforwardly extended to the general $\gamma$ case too.
In order to transform the cosmological system  (\ref{Fr1}),
(\ref{Fr2}) or (\ref{psievol}) and
(\ref{constraint2aa}) into its autonomous form, we introduce the
dimensionless variables  
\begin{equation}
x=\frac{\dot{\psi}}{\sqrt{6}H},\, y=\frac{\sqrt{W(\psi)}}{\sqrt{3}H},\, 
u=\frac{a_{ref}}{a}, \, v=\frac{V(\psi)}{H^2}.\label{auxiliary}
\end{equation}
Taking   derivatives with respect to $\ln a$, we obtain  the autonomous
form of the cosmological system  as
{\small
\begin{eqnarray}
 &&x'= (q-2) x+\sqrt{\frac{3}{2}} \lambda  y^2 
\nonumber\\
 &&+  \frac{3 v
\left[3 \alpha_3+\alpha_4+u^2 (\alpha_3+\alpha_4)-2 u (2
\alpha_3+\alpha_4+1)+3\right] \left\{4 \alpha_3+\alpha_4+u \left[\alpha_3
(u-5)+\alpha_4 (u-2)-3\right]+6\right\}}{2 x \left[-3 \alpha_3
(u-1)+\alpha_4 (u-1)^2+3\right]}
\nonumber\\ 
&&
-\frac{3 B v \left[3
   \alpha_3+\alpha_4+u^2 (\alpha_3+\alpha_4)-2 u (2
\alpha_3+\alpha_4+1)+3\right]}{2 x}\label{autonomous1},
\end{eqnarray}
\begin{eqnarray}
 y'=y \left(q-\sqrt{\frac{3}{2}} \lambda  x+1\right)\label{autonomous2},
\end{eqnarray}
\begin{eqnarray}
u'=-u\label{autonomous3},
\end{eqnarray}
\begin{eqnarray}
&&v'=  \left\{(u-1) \left[-3 \alpha_3
(u-1)+\alpha_4 (u-1)^2+3\right]^2 \right\}^{-1}\left\{3 v \left(3
\alpha_3+\alpha_4+\alpha_4 u^2-3 \alpha_3 u-2 \alpha_4
u+3\right)\right.\nonumber\\
&&\left.\ \ \ \ \ \ \ \ \ \ \ \ \ \ \ \ \ \ \   \ \ \ \ \ \ \ \ \ \ \
\times
 \left[3
\alpha_3+\alpha_4+u^2 (\alpha_3+\alpha_4)-2 u (2
\alpha_3+\alpha_4+1)+3\right]\right\}   
+   2 (q+1) v
.\label{autonomous4}
\end{eqnarray}}

Furthermore, using  \eqref{Fr1},\eqref{Fr2} and (\ref{wdedef}) 
we can
express the dark energy density parameter, the  dark
energy equation-of-state parameter  and the  deceleration parameter,  in
terms of the auxiliary
variables respectively as 
 \begin{equation} 
  \Omega_{DE}=\frac{1}{3} \left\{(u-1) v \left\{u \left[(u-5)
\alpha_3+(u-2)
\alpha_4-3\right]+4
\alpha_3+\alpha_4+6\right\}+3 \left(x^2+y^2\right)\right\},   
\end{equation}
  \begin{eqnarray} 
&&w_{DE}=\frac{v \left[4 \alpha_3+\alpha_4+u^2 (2 \alpha_3+\alpha_4+1)-2 u
(3 \alpha_3+\alpha_4+3)+6\right]+x^2-y^2}{(u-1) v \left\{4
   \alpha_3+\alpha_4+u \left[\alpha_3 (u-5)+\alpha_4
(u-2)-3\right]+6\right\}+x^2+y^2}\nonumber 
\\
 &&\ \ \ \ \ \ \ \ \ \ -\frac{B v \left[3 \alpha_3+\alpha_4+u^2
(\alpha_3+\alpha_4)-2 u (2 \alpha_3+\alpha_4+1)+3\right]}{(u-1) v \left\{4
   \alpha_3+\alpha_4+u \left[\alpha_3 (u-5)+\alpha_4
(u-2)-3\right]+6\right\}+x^2+y^2},
\end{eqnarray}
  \begin{eqnarray}
   && q=\frac{1}{2} \left\{3 v \left[4 \alpha_3+\alpha_4+u^2 (2
\alpha_3+\alpha_4+1)-2 u (3 \alpha_3+\alpha_4+3)+6\right]+3 x^2-3
   y^2+1\right\}
\nonumber\\ 
&&  \ \ \ \ \ -\frac{3}{2} B v \left[3 \alpha_3+\alpha_4+u^2
(\alpha_3+\alpha_4)-2 u (2 \alpha_3+\alpha_4+1)+3\right].
   \end{eqnarray}

In summary, (\ref{autonomous1})-(\ref{autonomous4}) account for an autonomous system 
defined in the physical phase space given by
{\small
\begin{align}
&\left\{(x,y,u,v): 
0\leq\frac{1}{3} \left\{(u-1) v \left\{u \left[(u-5) \alpha_3+(u-2)
\alpha_4-3\right]+4
\alpha_3+\alpha_4+6\right\}+3 \left(x^2+y^2\right)\right\}\leq 1, \right.
\nonumber \\  & \ \ \ \ \ \  \ \ \ \ \ \ \ \ \  \ \ \ \left . \frac{(u-1) v
\left(u^2 \alpha_4-3 u \alpha_3-2 u
\alpha_4+3 \alpha_3+\alpha_4+3\right)}{u^3}<0, u\geq 0, v\geq 0\right\},
\end{align}
}
where the first inequality follows from the physical condition $0 \leq 
\Omega_{DE} \leq 1,$ and the second inequality follows from the requirement the graviton
mass square $V(\psi)$ to remain positive.

 \begin{table*}[!]
\begin{center}
\begin{tabular}{|c|c|c|c|c|c|}
\hline
&&&& &  \\
 Cr. P.& $x_c$ & $y_c$ & $u_c$ & $v_c$& Exists for \\
\hline \hline
$R_1$& 0 & 0 & 0  & 0 &  all $\lambda_W$
\\[0.2cm]
\hline
 $R_2$& 1& 0 & 0 & 0 & all $\lambda_W$ \\[0.2cm]
\hline
 $R_3$& -1& 0 & 0 &  0 &  all $\lambda_W$ \\[0.2cm]
\hline
$R_4^\pm$& $\frac{\lambda_W}{\sqrt{6}}$ &
$\pm\sqrt{1-\frac{\lambda_W^2}{6}}$ &
0
& 0 &   $ 0<\lambda_W^2\leq6$ \\[0.2cm]
\hline
 $R_5^\pm$& $\sqrt{\frac{3}{2}}\frac{1}{\lambda_W}$ &
$\pm\sqrt{\frac{3}{2\lambda_W^2}}$ &
0 &  0 &    $\lambda_W^2\geq3$ \\[0.2cm]
\hline
$R_6^\pm$&0 & $\pm1$ & 0& 0 &   $\lambda_W=0$ \\[0.2cm]
\hline
$R_7$& $\sqrt{\frac{3}{2}}\frac{1}{\lambda_W}$ &
$y_c$ & 0&
$\frac{2 \lambda_W^2 y_c^2-3}{2 \lambda_W ^2 \mu_1}$ &   $\lambda_W\neq0$,
$\frac{2 \lambda ^2 y_c^2-3}{\mu_1}\geq 0$ \\[0.2cm]
&  & & & &   $0\leq 2 (1-\mu_2) \lambda_W ^2 y_c^2+3 (1+\mu_2)\leq 2
\lambda_W ^2$ \\[0.2cm]
\hline
$R_8$ & $x_c$ & $0$ & $0$ & $-\frac{x_c^2}{\mu_1}$ & $0\leq
x_c^2(1+\mu_2)\leq 1$ \\[0.2cm] \hline
\end{tabular}
\end{center}
\caption[crit]{\label{crit} The real and physically meaningful
critical points of the autonomous system
(\ref{autonomous1})-(\ref{autonomous4})
and their existence conditions. We have
introduced the notations    
$\mu_1=\left[4 \alpha_3+\alpha_4-B (3 \alpha_3+\alpha_4+3)+6\right]$ and
$\mu_2=\frac{4 \alpha_3+\alpha_4+6}{4 \alpha_3+\alpha_4-B (3
\alpha_3+\alpha_4+3)+6}.$}
\end{table*}

\begin{table*}[!]
\begin{center}
\begin{tabular}{|c|c|c|c|c|}
\hline
 & &&&  \\
 Cr. P.  &
Stability  &$\Omega_{DE}$ &$w_{DE}$ &$q$\\
\hline \hline
$R_1$ & saddle point  &0&arbitrary&$\frac{1}{2}$    \\
\hline
 $R_2$  &     saddle point  &1&1&2\\
\hline
 $R_3$  &     saddle point   &1&1&2\\
\hline
$R_4^\pm$   & stable
node for $0<\lambda_W^2<3$  & 1 &
$-1+\frac{\lambda_W^2}{3}$ &  
$-1+\frac{\lambda_W^2}{2}$\\
&   saddle point for
 $3<\lambda_W^2<6$  &&&\\
\hline
 $R_5^\pm$  & non-hyperbolic  &
$\frac{3}{\lambda_W^2}$ &  0 &
$\frac{1}{2}$\\
&   3D stable manifold for  &&&\\
&    $3<\lambda_W^2<\frac{24}{7}$ (stable node)
 &&&\\
&  or  $\lambda_W^2>\frac{24}{7}$ (stable spiral) &&&\\
\hline
$R_6^\pm$  & stable node  &1&$-1$&$-1$\\
\hline
$R_7$  & stable   & $\frac{3 (\mu_2+1)}{2 \lambda_W ^2}+(1-\mu_2)
y_c^2$
   & 0     &  $\frac{1}{2}$\\
\hline
$R_8$  & stable for
$x_c \lambda_W>\sqrt{\frac{3}{2}}$   & $x_c^2(1+\mu_2)$     & 0  & 
$\frac{1}{2}$
\\
&  saddle point otherwise&&& \\
\hline
\end{tabular}
\end{center}
\caption[crit]{\label{Tab3}
The stability conditions and the values of the observables $\Omega_{DE}$, 
$w_{DE}$ and $q$, for the real and physically meaningful  critical points 
of the
autonomous system (\ref{autonomous1})-(\ref{autonomous4}). The notations
are the same with Table \ref{crit}. }
\end{table*}

The real and physically meaningful critical points  $(x_c,y_c,u_c,v_c)$ of
the autonomous system  (\ref{autonomous1})-(\ref{autonomous4}), along with
their existence conditions, are presented  in Table \ref{crit}.
For each critical point  we calculate the $4\times4$
matrix ${\bf {Q}}$ of the linearized perturbation equations, and we
determine its type and stability by examining the sign of the real part of
the eigenvalues of ${\bf {Q}}$. The details of the analysis and the various
eigenvalues are presented in Appendix \ref{App3}, and in Table \ref{Tab3}
we display the stability conditions and the corresponding values of the
observables $\Omega_{DE}$, 
$w_{DE}$ and $q$.

 We mention here  that the variable choice
(\ref{auxiliary}) allows for an easy   classification of
expanding and contracting solutions. In particular, solutions with  
$y>0$ correspond to $H>0$ and thus to expansion, while those
with $y<0$ correspond to $H<0$ and therefore to contraction. That is why
points with
$y>0$ are denoted with the subscript ``+'', while those with
$y<0$ are denoted with the subscript ``-''.
However,   from  \eqref{autonomous2} it is implied
that the sign of $y$ is invariant, and thus    
transitions from contracting to expanding solutions or vice
versa do not exist (there could still exist  at the edge of the phase
space, which could be revealed only through application of Poincar\'e
central projection method \cite{PoincareProj,Leon2011,Leon:2012mt}, but
such an analysis lies beyond the scope of the present work and it is left
for
future investigation).

\subsubsection{Open universe}

In this case it proves convenient to impose the ansatz $b(t)=b_0 a(t)$,
since this leads to $\dot{b}=b_0 \dot{a}$, which
simplifies significantly the cosmological equations. In the following, we
focus on the dust matter ($\gamma=1$), however the analysis can be
straightforwardly extended to the general $\gamma$ case too.
In order to transform the cosmological system  (\ref{Fr1b}), (\ref{Fr2b})
or
(\ref{psievolb}) and (\ref{constraintb}) into its autonomous form, we
introduce the
dimensionless variables 
\begin{equation}
x=\frac{\dot{\psi}}{\sqrt{6}H},\, y=\frac{\sqrt{W(\psi)}}{\sqrt{3}H},\, 
u=\frac{k  }{a},\, v=\frac{V(\psi)}{H^2},\, \Omega_k=\frac{ k  }{a
H},\label{auxiliaryb}
\end{equation} where $k=\sqrt{\left|K\right|}.$
Taking   derivatives with respect to $\ln a$ we obtain  the autonomous
form of the cosmological system as
\begin{eqnarray}
&&x'=\frac{3 u^2 v x \beta  \delta }{\Omega_k^3}-\frac{3 u^2 v x \delta 
\left[2
\alpha_3 \left(\beta ^2+\beta -2\right)-\alpha_4 (\beta +1) (\beta
   -1)^2-6\right]}{\Omega_k^2 \left[3 \alpha_3 (\beta -1)-\alpha_4 (\beta
-1)^2-3\right]}  -\frac{\Omega_k^2 x}{2}+\frac{3
x^3}{2}  \ \ \ \  \  \    \ \ \ \  \   
\nonumber\\ 
&&\ \ \ \ \ \  \ +\sqrt{\frac{3}{2}} y^2
\lambda_W+\frac{v x \delta }{2
   \Omega_k} \left\{-\frac{6 u^2 \left[\alpha_3
   (\beta -4) (\beta -1)+\alpha_4 (\beta -1)^2-3 \beta +6\right]}{3
\alpha_3 (\beta -1)-\alpha_4 (\beta -1)^2-3}-\beta \right\}
\nonumber\\ 
&&\ \ \ \ \ \  \ +\frac{1}{2} x \left\{v \left[\beta ^2 (2
\alpha_3+\alpha_4+1)-2 \beta  (3 \alpha_3+\alpha_4+3)+4
\alpha_3+\alpha_4+6\right]-3
\left(y^2+1\right)\right\},\ \ \ \ \
\label{autonomousb1}
	\end{eqnarray}
\begin{eqnarray}
   && y'=\frac{1}{2} y \left\{v
\left[\beta
^2 (2 \alpha_3+\alpha_4+1)-2 \beta  (3 \alpha_3+\alpha_4+3)+4
   \alpha_3+\alpha_4+6\right]-3 y^2+3\right\} \ \ \ \  \  \ \ \ \  \  \ \ \
\  \ \ \ 
\nonumber\\ 
&&\ \ \ \ \ \ \   -\frac{v y \beta  \delta }{2 \Omega_k}
-\frac{\Omega_k^2 y}{2}+\frac{3 x^2
y}{2}-\sqrt{\frac{3}{2}} x y \lambda_W, \label{autonomousb2}
\\
   &&u'=-u, \label{autonomousb3}
	\end{eqnarray}
\begin{eqnarray}
   && v'=-\frac{v^2 \beta  \delta }{\Omega_k}-v \Omega_k^2+3 v
x^2+\frac{18 u^2 v x^2
\delta }{\Omega_k^2 (\beta -1) \left[-3 \alpha_3 (\beta -1)+\alpha_4 (\beta
-1)^2+3\right]}
\nonumber\\ 
&&\ \ \ \ \ \  \ +v \left\{v \left[\beta
^2 (2
   \alpha_3+\alpha_4+1)-2 \beta  (3 \alpha_3+\alpha_4+3)+4
\alpha_3+\alpha_4+6\right]-3 y^2+3\right\}
\nonumber\\
&&\ \ \ \ \ \  \ -\frac{18 u^2 v x^2 \delta }{\Omega_k (\beta -1)
   \left[-3 \alpha_3 (\beta -1)+\alpha_4 (\beta
-1)^2+3\right]},\label{autonomousb4}
\\
  && \Omega_k'=\Omega_k \left\{\frac{1}{2} \left\{v \left[\beta ^2 (2
\alpha_3+\alpha_4+1)-2 \beta  (3 \alpha_3+\alpha_4+3)+4
\alpha_3+\alpha_4+6\right]-3 y^2+1\right\}+\frac{3
x^2}{2}\right\} \nonumber\\ 
&&\ \ \ \ \ \  \ 
 -\frac{v
\beta  \delta }{2}-\frac{\Omega_k^3}{2},
\label{autonomousb5}
\end{eqnarray}
where $\beta=b_0 k$ and $\delta=\beta ^2 (\alpha_3+\alpha_4)-2 \beta  (2
\alpha_3+\alpha_4+1)+3 \alpha_3+\alpha_4+3.$

Furthermore, using (\ref{Fr1b}),(\ref{Fr2b})
we can
express the dark energy density parameter, the  dark
energy equation-of-state parameter  and the  deceleration parameter,  in
terms of the auxiliary
variables respectively as
\begin{eqnarray} 
&& \Omega_{DE}=\frac{1}{3} \left\{v (\beta -1) \left[\alpha_3 (\beta
-4) (\beta -1)+\alpha_4 (\beta -1)^2-3 \beta +6\right]+3
   \left(x^2+y^2\right)\right\},\nonumber\\
&&w_{DE}= \frac{v \left[\beta ^2 (2 \alpha_3+\alpha_4+1)-2 \beta  (3
\alpha_3+\alpha_4+3)+4 \alpha_3+\alpha_4+6\right]+3
   \left(x^2-y^2\right)}{(\beta -1) v \left[\alpha_3 (\beta -4) (\beta
-1)+\alpha_4 (\beta -1)^2-3 \beta +6\right]+3
   \left(x^2+y^2\right)} \nonumber\\ 
&&\ \ \ \ \ \ \ \ \ \  -\frac{\beta  v \left[\beta ^2
(\alpha_3+\alpha_4)-2 \beta  (2 \alpha_3+\alpha_4+1)+3
\alpha_3+\alpha_4+3\right]}{\Omega_k \left\{(\beta -1) v \left[\alpha_3
(\beta -4) (\beta -1)+\alpha_4 (\beta -1)^2-3 \beta +6\right]+3
   \left(x^2+y^2\right)\right\}} ,\nonumber\\
   && q=-\frac{\beta  v \left[\beta ^2 (\alpha_3+\alpha_4)-2 \beta  (2
\alpha_3+\alpha_4+1)+3 \alpha_3+\alpha_4+3\right]}{2 \Omega_k}
-\frac{\Omega_k^2}{2}
\nonumber \\ 
&&\ \ \ \ \, \   +\frac{1}{2} \left\{v \left[\beta ^2 (2
\alpha_3+\alpha_4+1)-2 \beta  (3 \alpha_3+\alpha_4+3)+4
   \alpha_3+\alpha_4+6\right]+3 x^2-3 y^2+1\right\}.\ \ \ \ \ \  \ \ \  
   \end{eqnarray}

In summary, the autonomous system
(\ref{autonomousb1})-(\ref{autonomousb5}) defines a flow  in the physical
phase space  given by
\begin{eqnarray}
&&\left\{(x,y,u,v,\Omega_k): 0\leq \frac{1}{3} \left\{v (\beta -1)
\left[\alpha_3 (\beta
-4) (\beta -1)+\alpha_4 (\beta -1)^2-3 \beta
+6\right]\right.\right.\nonumber\\
&&\ \ \ \ \ \  \  \ \ \ \ \  \  \ \ \ \ \ \  \  \ \ \ \ \  \  \ \ \ \ \  \
\left.\left.+3
   \left(x^2+y^2\right)\right\}+\Omega_k^2\leq 1, u\geq 0, v\geq 0\right\}.
\end{eqnarray}
as it arises from the physicality requirements $a\geq0$, $V(\psi)\geq0$,
$W(\psi)\geq0$ and $0\leq\Omega_{DE}+\Omega_k^2\leq1.$ 

 \begin{table*}[!]
\begin{center}
\begin{tabular}{|c|c|c|c|c|c|c|}
\hline
&&&&& &  \\
 Cr. P.& $x_c$ & $y_c$ & $u_c$ & $v_c$ & $\Omega_{kc}$ & Exists for \\
\hline \hline
$S_1$& 0 & 0 & 0  & 0 & 0 &  always
\\
\hline
 $S_2$& 1& 0 & 0 & 0 & 0&  always \\
\hline
 $S_3$& -1& 0 & 0 &  0 & 0&  always \\
\hline
$S_4^\pm$& $\frac{\lambda_W}{\sqrt{6}}$ &
$\pm\sqrt{1-\frac{\lambda_W^2}{6}}$ &
0
& 0 & 0&   $ \lambda_W^2\leq6$ \\
\hline
 $S_5^\pm$& $\sqrt{\frac{3}{2}}\frac{1}{\lambda_W}$ &
$\pm\sqrt{\frac{3}{2\lambda_W^2}}$ &
0 &  0 & 0&    $\lambda_W^2\geq3$ \\
\hline
$S_6^\pm$&0 & $\pm1$ & 0& 0 &  0&  $\lambda_W=0$ \\
\hline
$S_7^\pm$& 0 & 0 & 0& 0 & $\pm1$&   always \\
\hline
$S_8^\pm$& $\sqrt{\frac{3}{2}}\frac{1}{\lambda_W}$ &
$\pm\frac{2}{\sqrt{3\lambda_W^2}}$
& 0& 0 & $\pm\sqrt{1-\frac{2}{\lambda_W^2}}$&  $\lambda_W^2\geq 2$  \\
\hline
\end{tabular}
\end{center}
\caption[crit]{\label{critb}
The real and physically meaningful
critical points of the autonomous system
(\ref{autonomousb1})-(\ref{autonomousb5})
and their existence conditions. }
\end{table*}

 \begin{table*}[!]
\begin{center}
\begin{tabular}{|c|c|c|c|c|}
\hline
&&&&    \\
 Cr. P. & Stability & $\Omega_{DE}$ &  $w_{DE}$ &  $q$\\
\hline \hline
$S_1$&saddle point & 0 & arbitrary   &$\frac{1}{2}$ \\
\hline
 $S_2$&saddle point &
1  &   1  & 2\\
\hline
 $S_3$&saddle point &
1  &   1   & 2\\
\hline
$S_4^\pm$  &saddle point &  1 &
$-1+\frac{\lambda_W^2}{3}$ &  
$-1+\frac{\lambda_W^2}{2}$\\
\hline
 $S_5^\pm$&saddle point & $\frac{3}{\lambda_W^2}$ &   0 &
$\frac{1}{2}$\\
\hline
$S_6^\pm$& non-hyperbolic (4D stable manifold) & 1 &  $-1$ & $-1$\\
\hline
$S_7^\pm$&saddle point& 0    & arbitrary     &  0\\
\hline
$S_8^\pm$ &saddle point& $\frac{2}{\lambda_W ^2}$   &
$-\frac{1}{3}$ & 0\\
\hline
\end{tabular}
\end{center}
\caption[crit]{\label{crit3}  
The stability conditions and the values of the observables $\Omega_{DE}$, 
$w_{DE}$ and $q$, for the real and physically meaningful  critical points 
of the
autonomous system (\ref{autonomousb1})-(\ref{autonomousb5}).   }
\end{table*}

The real and physically meaningful critical points 
$(x_c,y_c,u_c,v_c,\Omega_{kc})$ of
the autonomous system  (\ref{autonomousb1})-(\ref{autonomousb5}), along
with
their existence conditions, are displayed  in Table \ref{critb}.
For each critical point   we calculate the $5\times5$
matrix ${\bf {Q}}$ of the linearized perturbation equations, and we
determine its type and stability by examining the sign of the real part of
the eigenvalues of ${\bf {Q}}$. The details of the analysis and the various
eigenvalues are presented in Appendix \ref{App4}, and in Table \ref{crit3}
we display the stability conditions and the corresponding values of the
observables $\Omega_{DE}$, 
$w_{DE}$ and $q$.

Note that the variable choice
(\ref{auxiliaryb}) allows for an easy classification of
expanding and contracting solutions. In particular, solutions with  
$\Omega_k=k/(aH)>0$ or $y>0$ correspond to $H>0$ and thus to expansion,
while those
with $\Omega_k<0$ or $y<0$ correspond to $H<0$ and therefore to
contraction.  However, from the equations \eqref{autonomousb2} and
\eqref{autonomousb5} we deduce
that the sign of   $y$ and the sign of $\Omega_k$ are invariant and thus 
transitions from contracting to expanding solutions or vice
versa do not exist. 
Nevertheless, since such transitions do exist in the
flat geometry \cite{Cai:2012ag}, there could still exist in the non-flat
scenario too, at the edge of the phase space, which
could be revealed only through application of Poincar\'e
central projection method \cite{PoincareProj,Leon2011,Leon:2012mt}. This
analysis lies beyond the scope of the present work and it is left for
a future project.

\section{Cosmological Implications}
\label{implications}

In the previous section we performed a complete dynamical analysis of the
scenario of extended (varying mass) nonlinear massive gravity for both flat
and open FRW geometries, we extracted the late-time stable solutions and we
calculated the corresponding observables. In this section we  
discuss the cosmological implications of the various scenarios case by
case.

\subsection{Imposing $V(\psi)$ at will}

\subsubsection{Flat universe}

First of all we mention that the scenario at hand coincides with standard
quintessence if the graviton mass square $V(\psi)=V_0 e^{-\lambda_V\psi}$
is
identically zero. If this is not the case then standard quintessence can
be obtained only asymptotically. Additionally, if $V_0$ is not zero then
$\lambda_V$ cannot be
zero, since the constraint (\ref{constraint})
cannot be satisfied in general. Thus, we conclude that in general this
scenario has $\lambda_V\neq0$, and therefore there are not parameter
values that make it coincide completely with 
usual (constant mass) massive gravity, as discussed in
\cite{Saridakis:2012jy}.

As we observe from Table \ref{Tab1},  there exist three
critical points
and all of them can be stable according to the parameter values.
Point $P_1$  in the case of standard matter
($\gamma=1$)   corresponds to a non-accelerating 
universe, with a   dark energy behaving as dust. Although it has the
advantage that $0<\Omega_{DE}<1$, that is it can alleviate
the coincidence problem since dark energy and dark matter density
parameters can be of the same order, the above features disfavor it.
Lastly, note
that the corresponding graviton mass has become zero.

Point $P_2$ corresponds to a dark-energy dominated, non-accelerating, 
universe,  with   dark-energy behaving as dust, and thus it is also
disfavored by observations. Additionally, the graviton mass remains
finite.

Point $P_3$ is the most interesting solution that can attract the universe
at late times. It corresponds to a dark-energy dominated universe,
which can be accelerating (for $\frac{\lambda_W}{\lambda_V}<\frac{2}{3}$)
or non-accelerating according to the parameter values, and where  dark
energy can lie either in the quintessence \cite{quint,quint1} or in the
phantom regime \cite{Caldwell:1999ew} (for
$\lambda_W\lambda_V<0$). Moreover, the graviton mass dynamically  becomes
 zero. These features make this point a very good candidate for
the description of late-time universe, in agreement with observations.
Furthermore, note that if the universe starts from the quintessence
regime, then the attraction to $P_3$  implies the
phantom-divide crossing \cite{Cai:2009zp}.  The
realization of the phantom regime and/or of the phantom-divide crossing,
is a  great advantage of extended
nonlinear massive gravity, as was analyzed in detail in
\cite{Saridakis:2012jy,Cai:2012ag}.
 \begin{figure}[ht]
\begin{center}
\includegraphics[height=8cm,width=8cm]{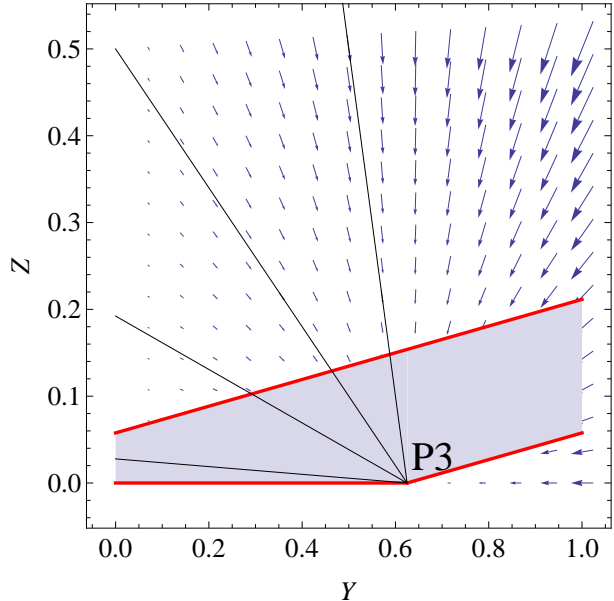}
\caption{\label{fig1}{\it{ Trajectories in the $Y$-$Z$ plane  of the
cosmological scenario \eqref{ODEs1}, where the varying graviton mass square
$V(\psi)$ is imposed at will, in a flat universe. We use  $\gamma=1,
\lambda_V=2,\lambda_W=-1, \alpha_3=\alpha_4=0.1$. The physical part of the
phase space is marked by the shadowed region limited by the red lines. In
this specific example the universe is led to the phantom
stable late-time solution $P_3$.}} } 
\end{center}
\end{figure}

 We mention here that naively it
looks strange that $P_3$ can be a phantom solution although the graviton
mass tends to zero and the model should look like quintessence. However,
this is easily explained since, as we show in Appendix \ref{App1}, in the
case $\lambda_W\lambda_V<0$ where $P_3$ is phantom, $V$ tends to zero but
$W$ and $H$ tend to infinity, which is a Big-Rip-type behavior (realized
at infinity and not at a finite scale factor)
\cite{Sami:2003xv,Nojiri:2005sx,Copeland:2006wr,Briscese:2006xu,
Bamba:2008ut,Capozziello:2009hc,
Saridakis:2009jq}, that is a typical fate of phantom scenarios. In other
words, the graviton mass does tend asymptotically to zero, but its
previous effect has already led the universe to a phantom regime without
exit (although not so strongly in order to exhibit a Big Rip at a finite
scale factor).

In order to present the above behavior in a more transparent way, we evolve
numerically the autonomous system \eqref{ODEs1}  in the
invariant set $u=0$,   
 for the  the parameters $\gamma=1$ (dust matter),
$\lambda_V=2,\lambda_W=-1, \alpha_3=\alpha_4=0.1$, and in Fig. \ref{fig1}
we depict the corresponding phase-space behavior  in the 
$Y$-$Z$ plane. The physical part of the
phase space is marked by the shadowed region limited by the red lines. As
we
observe, in this specific example the universe results in the phantom
stable late-time solution $P_3$.

\subsubsection{Open universe}

As we show  in detail in Appendix \ref{App2}, and as we have depicted in
Tables
\ref{crit01open1}, \ref{crit01open2} and \ref{quintessence},   the
scenario at hand admits many
stable late-time solutions, and this reveals its advantages and
capabilities, comparing to standard quintessence, as well as to standard
(constant-mass) nonlinear massive gravity. Note that this scenario admits
also curves of solutions apart from individual points, which is an
additional indication of its generalized features.

In particular, the first interesting solution, that can attract the
universe at late times, is the curve of critical points $Q_1$. It
corresponds to  an accelerating universe, in which dark energy  lies
always in the phantom regime. Moreover, it has the advantage that
$0<\Omega_{DE}<1$, that is it can alleviate the coincidence problem, and
the graviton mass can be zero or not according to the parameter values.

The curves of critical points $Q_3$ and $Q_6^\pm$, as well as the
individual critical
points $Q_7^\pm$, can be stable and thus attract the universe at
late
times, however since they correspond to zero acceleration are not favored
by observations (although they have $0<\Omega_{DE}<1$ and thus they can
solve the coincidence problem).

The curves of critical points $Q_4$ and $Q_5$ can be stable
(although their stable manifold has smaller dimensionality and thus the
stability is weaker), that is they can be the late-time state of the
universe, corresponding to an accelerating or non-accelerating universe
according to the parameter values. Furthermore, note that according to the
parameter values they can lie in
the quintessence or phantom regime, and
they possess $0<\Omega_{DE}<1$. Additionally, the graviton mass becomes
zero. These features make $Q_4$ and $Q_5$ good candidates for the
description of the universe. 

In particular, as we discussed in paragraph \ref{Vgivenopen}, the curve
$Q_5$ contains the quintessence-like critical points presented in Table
\ref{quintessence}, which are obtained in standard quintessence too in flat
\cite{Copeland:1997et} or non-flat geometries \cite{Copeland:2009be}. Note
however that the stability properties are slightly different, since now we
have the additional direction of the graviton mass. Amongst these points,
  $Q_{14}$ is stable, corresponding to a dark energy-dominated,
quintessence universe, which can be accelerating or non-accelerating
according to the parameter values, and thus it is a good candidate for the
description of the universe. On the other hand point $Q_{15}$, which is
stable in standard quintessence, in the present case it becomes saddle
and therefore it cannot be the late-time state of the universe.
\begin{figure}[!]
\begin{center}
\includegraphics[height=7.97cm,width=7.97cm]{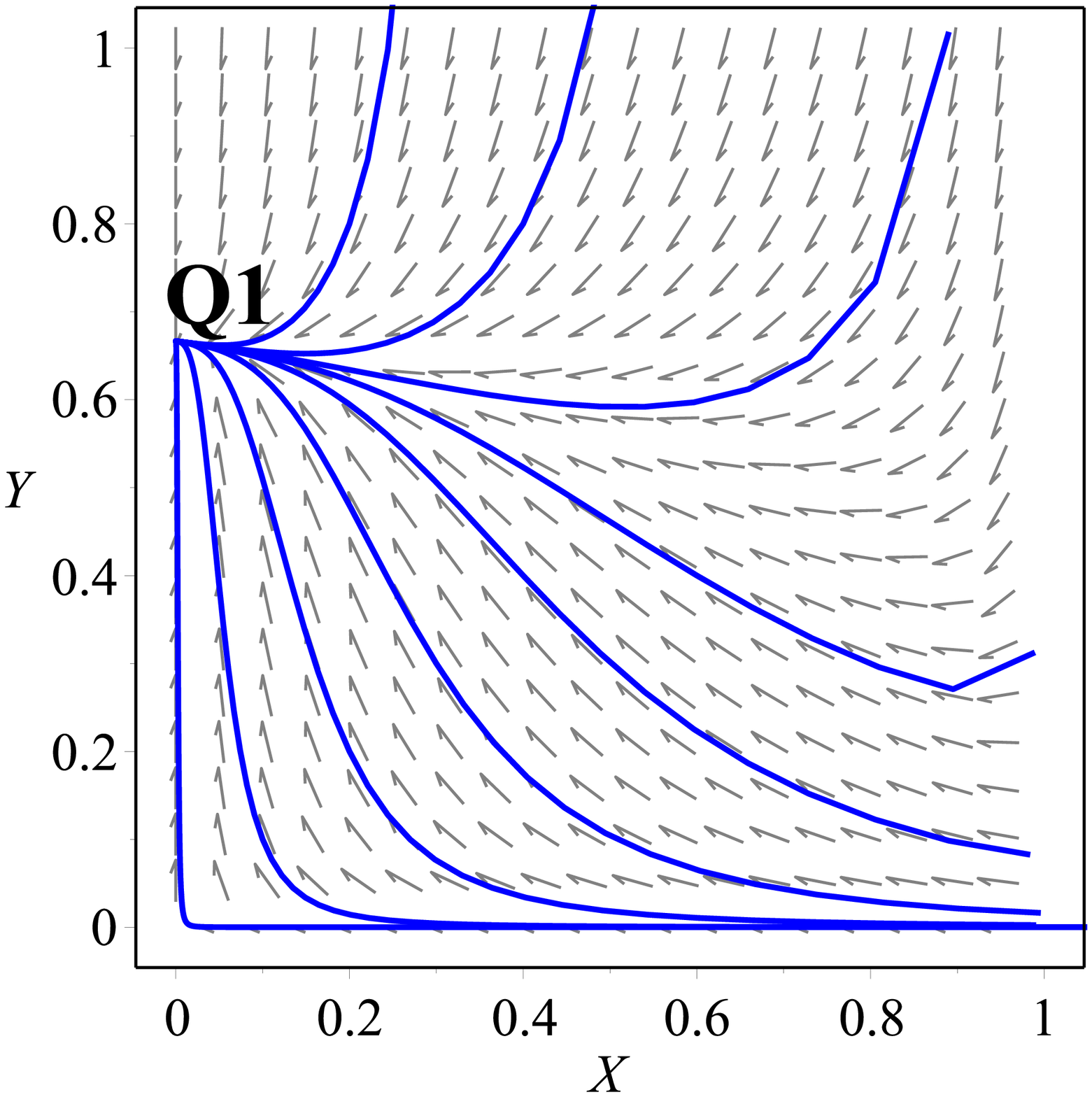}
\caption{\label{fig2}{\it{
Trajectories in the $X$-$Y$ plane  of the
cosmological scenario \eqref{ODEs2}, where the varying graviton mass square
$V(\psi)$ is imposed at will, in an open universe. We focus on the
invariant set
$\Omega_k=U=Z=0$ and we choose    $\gamma=1,
\lambda_V=-2,\lambda_W=1, \alpha_3=\alpha_4=0.1$.    In this specific
example the stable late-time state of the universe is the phantom
solution $Q_1$.  }} } 
\end{center}
\end{figure}
\begin{figure}[ht]
\begin{center}
\includegraphics[height=7.97cm,width=7.97cm]{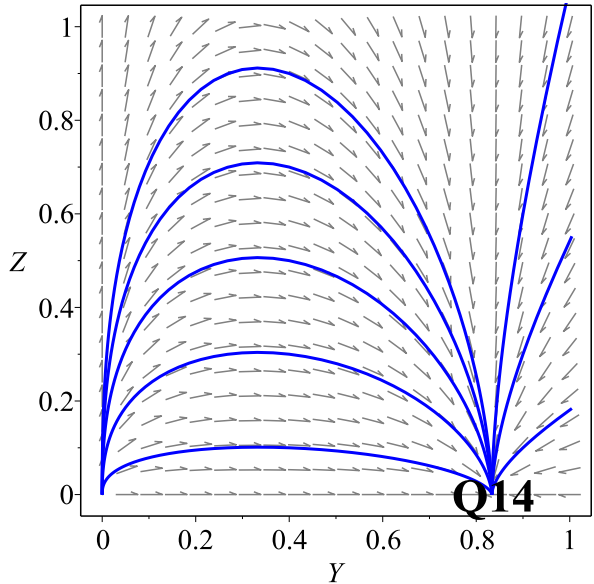}
\caption{\label{fig3}{\it{Trajectories in the $Y$-$Z$ plane  of the
cosmological scenario \eqref{ODEs2}, where the varying graviton mass square
$V(\psi)$ is imposed at will, in an open universe. We focus on the
invariant set
$\Omega_k=X=Z=0$ and we choose   $\gamma=1,
\lambda_V=2,\lambda_W=1, \alpha_3=\alpha_4=0.1$ and
$U_c=\frac{\lambda_V}{\sqrt{6}}$.    In this specific
example the stable late-time state of the universe is the quintessence-like
point $Q_{14}$.
  }} } 
\end{center}
\end{figure}
 \begin{figure}[!]
\begin{center}
\vspace{-3cm}
\includegraphics[scale=0.5]{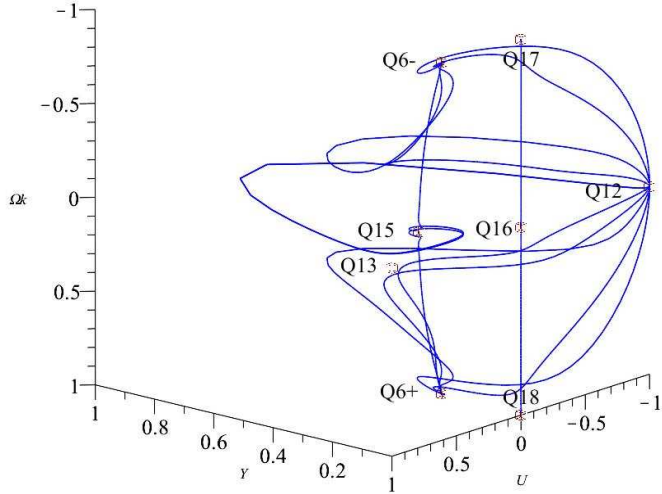}
\vspace{-3.cm}
\caption{\label{fig4}{\it{
Trajectories   of the
cosmological scenario \eqref{ODEs2}, where the varying graviton mass square
$V(\psi)$ is imposed at will, in an open universe, in the subset $X=Z=0$,
which is invariant provided $3+3\alpha_3+\alpha_4=0,\alpha_3\neq
-2,\alpha_4\neq 3$. We use the    parameters values
$\gamma=1,\lambda_W=3$. In this specific example the stable late-time
solutions of the universe are the expanding, non-accelerating $Q_6^+$ (its
basin of attraction is the half-subspace
$\Omega_k>0$), and the contracting  $Q_6^-$ (its basin of
attraction is the half-subspace
$\Omega_k<0$). Additionally, we can   see the saddle
points $Q_{15}$ (non-accelerating with $0<\Omega_{DE}<1$), $Q_{16}$ 
(non-accelerating, matter-dominated),  $Q_{17}$ (curvature-dominated,
contracting) and
$Q_{18}$ (non-accelerating, curvature-dominated, expanding), as well as
the unstable points $Q_{12}$ and $Q_{13}$  (non-accelerating, dark-energy
dominated, with stiff $w_{DE}$).
 }} } 
\end{center}
\end{figure}
Let us present the above results more transparently. In  Fig. \ref{fig2} we
show the corresponding phase-space behavior in the $X$-$Y$ plane, as it
arises from numerical elaboration of the autonomous system (\ref{ODEs2}). 
We focus on the invariant set $\Omega_k=U=Z=0$ and we choose    $\gamma=1,
\lambda_V=-2,\lambda_W=1, \alpha_3=\alpha_4=0.1$. In this specific
example the stable late-time state of the universe is the phantom
solution $Q_1$. 
Similarly, in Fig. \ref{fig3} we depict the corresponding phase-space
behavior of the   autonomous system (\ref{ODEs2}), but restricted to the
invariant set $\Omega_k=X=Z=0$, and using $\gamma=1,
\lambda_V=2,\lambda_W=1, \alpha_3=\alpha_4=0.1$ and
$U_c=\frac{\lambda_V}{\sqrt{6}}$. In this case the late-time stable
solution of the universe is the quintessence-like point $Q_{14}$.

Finally, in Fig. \ref{fig4} we present the phase-space behavior of the  
autonomous system (\ref{ODEs2}), in the subset $X=Z=0$, which is invariant
provided $3+3\alpha_3+\alpha_4=0$. In this case
the universe can be attracted by two stable late-time solutions, namely the
expanding, non-accelerating $Q_6^+$ (its basin of attraction is the
half-subspace $\Omega_k>0$), and the contracting  $Q_6^-$
(its basin of attraction is the half-subspace
$\Omega_k<0$).  Finally, as we discussed
in the end of paragraph \ref{Vgivenopen} and in Appendix \ref{App2}, we
mention that in the scenario at hand the sign of $\Omega_k$ is invariant.
Thus, although our model admits expanding (lower half of Fig. \ref{fig4})
and contracting evolution (upper half of Fig. \ref{fig4}),
there is no transition from contracting to expanding solutions or vice
versa, that is a cosmological bounce or turnaround is not possible.
 
In summary, as we can see, the scenario of extended nonlinear massive
gravity in open geometry has a great variety of stable late-time solutions,
as was shown in  \cite{Saridakis:2012jy,Cai:2012ag} through specific
examples.

\subsection{Imposing $b(t)$ at will}

\subsubsection{Flat universe}
 
In this case the scenario at hand admits a variety of stable late-time
solutions. In particular, point $R_4^+$ corresponds to an expanding
dark-energy
dominated universe, with dark energy  lying in the
quintessence regime, which can be accelerating or non-accelerating
according to the
usual potential exponent, and the graviton mass is zero. This point
exists in standard quintessence too \cite{Copeland:1997et}, and it is quite
important since it possesses $w_{DE}$ and $q$ compatible with
observations.

Point $R_5^+$ has the advantage that $0<\Omega_{DE}<1$, that is it can
alleviate the coincidence problem, and moreover the graviton mass is zero,
however it has the disadvantage that
it is not accelerating and possesses $w_{DE}=0$, which are not favored by
observations. This point exists in
standard quintessence too \cite{Copeland:1997et}, however note that in the
present case it is non-hyperbolic, and thus its stability is
weaker (due to the existence of an extra dimension in the phase space,
namely the graviton mass).

Point $R_6^+$ exists for $\lambda_W=0$ and it is always stable.
Although at first sight it seems
to be the $\lambda_W\rightarrow0$ limit of $R_4^+$   this
is not the case since the complete equations are different. It corresponds
to an accelerating, dark-energy dominated universe, in which dark energy
behaves like a cosmological constant, and moreover the graviton mass is
zero.
\begin{figure}[ht]
\begin{center}
\vspace{-4.8cm}
\includegraphics[scale=0.65]{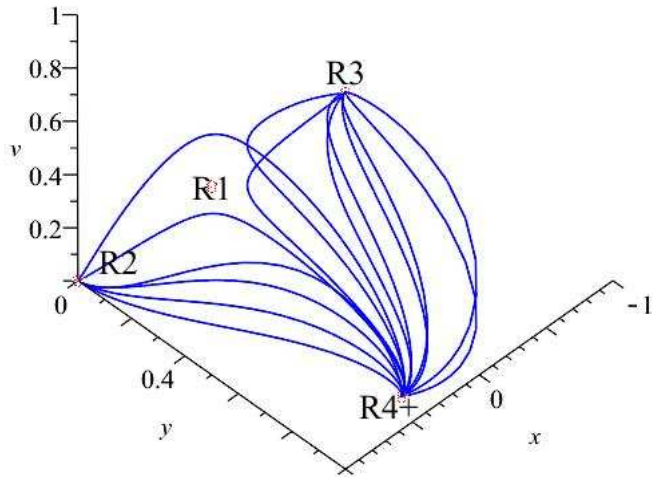}
\vspace{-4.5cm}
\caption{\label{fig5}{\it{
Trajectories   of the
cosmological scenario (\ref{autonomous1})-(\ref{autonomous4}), where the
St\"{u}ckelberg field function function $b(t)$
 is imposed at will, in a flat universe, using   
$\gamma=1,
\lambda_W=1, \alpha_3=\alpha_4=0.5, B=1.7$. 
In this specific example the stable late-time
state of the universe is the expanding, dark-energy
dominated, quintessence-like point  $R_4^+$.
Additionally, we depict the saddle
points $R_1$ (non-accelerating, matter-dominated), and $R_2$,$R_3$ 
(non-accelerating, dark-energy
dominated).
  }} } 
\end{center}
\end{figure}

The curves of critical points $R_7$ and $R_8$ can also be the late-time
state of the universe (they are non-hyperbolic and thus their stability
is weaker). They correspond to non-accelerating solutions, where the dark
energy behaves like dust and where $0<\Omega_{DE}<1$, and additionally
they
possess a non-zero value for the graviton mass. These features disfavor
these curves of critical points. Finally, we mention here that although
the aforementioned individual points were obtained in \cite{Wu:2013ii}
too, these curves of critical were missed, due to the fact that in the
analysis one of the phase-space directions was frozen for simplicity.

In   Fig. \ref{fig5} we depict orbits of the autonomous system
(\ref{autonomous1})-(\ref{autonomous4}), restricting to the
invariant set $u=0$, and using   
$\gamma=1,
\lambda_W=1, \alpha_3=\alpha_4=0.5, B=1.7$. 
In this specific
example the stable late-time state of the universe is the   expanding,
dark-energy
dominated, quintessence-like point  $R_4^+$.

\subsubsection{Open universe}

This scenario possesses only one stable solution that can attract the
universe at late times, namely $S_6^+$ (although at first sight it seems
to be the $\lambda_W\rightarrow0$ limit of $S_4^+$   this
is not the case since the complete equations are different). This point
corresponds to a
dark-energy dominated, accelerating universe, with zero graviton mass, and
where the dark energy behaves like cosmological constant. This solution  
  is the global attractor of this cosmological system, that is the
universe will be always led there, for every  initial conditions.   These
features make this point a good candidate
for the description of the universe. However, we mention that it
exists only for $\lambda_W=0$, that is a form of parameter-tuning is
needed. On the other hand, for $\lambda_W\neq0$ the system does not accept
\begin{figure}[ht]
\begin{center}
\vspace{-5.cm}
\includegraphics[scale=0.65]{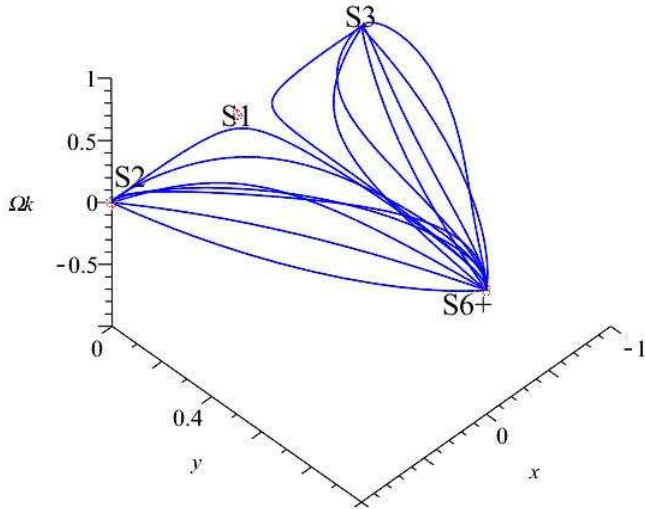}
\vspace{-5cm}
\caption{\label{fig6}{\it{
Trajectories   of the
cosmological scenario (\ref{autonomousb1})-(\ref{autonomousb5}), where the
St\"{u}ckelberg field function function $b(t)$
 is imposed at will, in a non-flat universe,   restricted to the
invariant set $u=v=0$, using   $\gamma=1$ and $\lambda_W=0$. In this
specific
example the stable late-time state of the universe is the 
cosmological-constant-like solution,  $S_6^+.$
Additionally, we depict the saddle
points $S_1$ (non-accelerating, matter-dominated), and $S_2$,$S_3$ 
(non-accelerating, dark-energy
dominated).
  }} } 
\end{center}
\end{figure}
any stable solutions, due to the fact that there are unstable directions
related to both curvature and graviton mass. In summary, this implies that
in general the scenario at hand has disadvantages, unless one tunes the
model parameters. Finally, note that since the sign of $\Omega_k$ is
invariant,   although the model admits expanding  
and contracting evolution,  a cosmological bounce or a turnaround is not
possible.

In   Fig. \ref{fig6} we present orbits of the autonomous system
(\ref{autonomousb1})-(\ref{autonomousb5}), restricting to the
invariant set $u=v=0$, and using   
$\gamma=1,
\lambda_W=0$. Note that the evolution is independent of the values of
$\alpha_3$, $\alpha_4$ and $b_0$, since
they do not appear explicitly in the equations governing the dynamics in
this invariant set.
In this specific
example the stable late-time state of the universe is the 
cosmological-constant-like solution,  $S_6^+.$

 \section{Conclusions}
\label{Conclusions}

In this work we   investigated the dynamical behavior of extended
(varying-mass) nonlinear massive gravity, which is an
extension of the usual nonlinear massive gravity  \cite{deRham:2010ik,
deRham:2010kj} where the graviton mass is promoted to a scalar-field
potential \cite{Huang:2012pe}. This scenario has a lot of freedom due to
the involved free functions, and thus its cosmological implications are
significant.

In order to extract the basic features of the above paradigm, we performed
a detailed dynamical analysis in the case of an open geometry, adding for
completeness the flat case, although it proves to have disadvantages that
can be cured only at the phenomenological level. In both analyses we
followed two approaches, namely the theoretically robust one to impose the
graviton mass square at will and let the equations determine suitably the
St\"{u}ckelberg field function, or the  theoretically less-justified one
to impose the St\"{u}ckelberg field function at will and let the equations
to determine the graviton mass square. In all cases we extracted the 
late-time solutions and  we calculated the corresponding observables, such
as the dark-energy
equation-of-state parameter, the deceleration parameter, and the
dark-energy and matter density parameter. 

One basic feature of the scenario at hand is that it can lead to an
accelerating universe, with an effective dark energy lying in the
quintessence or in the phantom regime, or experience the phantom-divide
crossing during the evolution. This is a
great advantage since the model at hand utilizes only a canonical field.
Additionally, and more interestingly, the universe cannot only be
phantom at one stage of its evolution, but also at its final
late-time solutions it can be quintessence or phantom like. This is not
the case in other modified-gravity scenarios, where the universe results
to quintessence-like solutions even if it has  passed through the phantom
regime \cite{Xu:2012jf}. The above features were discussed in
\cite{Saridakis:2012jy,Cai:2012ag} using specific solutions, but in the
present work they arise from a general dynamical analysis.

An additional advantage of extended nonlinear massive gravity  is that the
graviton mass goes asymptotically to zero at late times, without
fine-tuning, which is in agreement with observations. Note that this is not
the case in usual massive gravity, where ones needs to fine-tune the
graviton mass to a very small value by hand. 

Finally, another advantage of the present scenario is that the dark energy
density parameter at the late-time solutions can be between zero and one,
which can alleviate the coincidence problem since dark energy and dark
matter density parameters can be of the same order.

In the above analysis we used the exponential ansatz for the usual
scalar-field potential, and then we used an exponential form for the
graviton square mass, in order to be phenomenologically consistent.   
One could ask whether the above behaviors are a result of these specific
ansatzes, or they have a general character. Although this would need an
explicit investigation from the beginning, the details of our analysis
indicate 
that the results are qualitatively robust for many phenomenologically
consistent  varying graviton mass choices too. However, in the alternative
and less-justified approach where the  St\"{u}ckelberg field is imposed at
will, our results are quite sensitive to the input ansatz, and therefore a
detailed analysis is required for every new choice. The fact that the
results are very  sensitive in the St\"{u}ckelberg field ansatz, is known
to happen in the usual nonlinear massive gravity too
\cite{DeFelice:2012mx,D'Amico:2011jj,Gumrukcuoglu:2012aa}.

In summary, the scenario of extended (varying-mass) nonlinear massive
gravity, exhibits a larger variety and a richer structure of interesting
cosmological late-time solutions, comparing to usual quintessence,
phantom, and quintom cosmology, and also to   usual
(constant-mass) massive gravity. These features are in agreement with
observations and thus they make this paradigm a good candidate for the
description of nature. However, an additional requirement for the validity
of this scenario is to behave consistently beyond the background level too.
Since
the theory at hand is based on the usual massive gravity formalism
in order to become Boulware-Deser ghost free, the perturbation analysis
could reveal interesting issues too \cite{Deser:2012qx,Deser:2013uy}.
Although such a
perturbation investigation is therefore necessary, it lies beyond the scope
of the present work and it is left for a future project.

\begin{acknowledgments}
The authors would like to thank   S. Lepe for reading the original
manuscript and making helpful suggestions.
GL was supported by MECESUP FSM0806 from Ministerio de Educaci\'on, Chile 
and  by PUCV through Proyecto DI Postdoctorado 2013.
JS has been supported by Comisi\'on Nacional de Ciencias y Tecnolog\'ia
through
FONDECYT Grant 1110076, 1090613 and 1110230 and also by PUCV grant No.
123.713/2012.
The research project is implemented within the framework of the Action
«Supporting Postdoctoral Researchers» of the Operational Program
``Education and Lifelong Learning'' (Action’s Beneficiary: General
Secretariat for Research and Technology), and is co-financed by the
European Social Fund (ESF) and the Greek State.
\end{acknowledgments}

\begin{appendix}

\section{Stability when $V(\psi)$ is imposed at will}

\subsection{Flat universe}
\label{App1}

For the critical points $(u_c,Y_c,Z_c)$ of the autonomous  
system  (\ref{ODEs1}), the coefficients of the perturbation
equations form a $3\times3$ matrix ${\bf {Q}}$, however since they are
quite complicated expressions 
  we do not display them explicitly. Despite this complicated form, using
the specific critical points presented in Table \ref{Tab1}, the matrix
${\bf {Q}}$ obtains a simple form that allows for an easy calculation of
its eigenvalues. The corresponding eigenvalues  and the stability
conditions
for each critical point are presented in Table
\ref{Tab2b}.
 \begin{table*}[ht]
\begin{center}
\begin{tabular}{|c|c|c|}
\hline
&&   \\
 Cr. P. & Eingenvalues & Stability \\
\hline \hline
$P_1$&   $-1,3 (\gamma -1),3 \left(\gamma -\frac{
\lambda_W}{\lambda_V}\right)$ & stable for
$\gamma<\min\left\{1,\frac{\lambda_W}{\lambda_V}\right\},\lambda_V^2\geq
\frac{3}{2}$ 
\\[0.2cm]
&& saddle point otherwise  \\ \hline
$P_2$& $-1,-3 (\gamma -1),3 \left(1 -\frac{
\lambda_W}{\lambda_V}\right)$ & stable for
$\gamma>1,\frac{\lambda_W}{\lambda_V}>1$ 
\\[0.2cm]
&& saddle point otherwise  \\ \hline
$P_3$ & $-1,-3 \left(\gamma -\frac{
\lambda_W}{\lambda_V}\right) ,-3 \left(1 -\frac{
\lambda_W}{\lambda_V}\right)$ & $\frac{\lambda_W}{\lambda_V}<\min
\left\{1,\gamma\right\}, \lambda_V^2\geq \frac{3}{2}$  
\\[0.2cm]
&& saddle point otherwise  \\ 
 \hline
\end{tabular}
\end{center}
\caption[crit]{\label{Tab2b} The eigenvalues of   matrix
${\bf {Q}}$ of the perturbation equations of the autonomous system
(\ref{ODEs1}), and the corresponding stability conditions.}
\end{table*}

Since in the special case $\gamma=1$ (dust matter) one eigenvalue of $P_1$
and $P_2$ becomes zero, we need to examine this case separately. For
$\gamma=1$ the system \eqref{ODEs1} is restricted to the
invariant set $u=0$ and it admits the general solution 
\begin{align}
Y(\tau)=\frac{3-2 \lambda_V^2}{e^{c_1 \left(2
\lambda_V^2-3\right)-\frac{\lambda_W \tau }{\lambda_V}+\tau }-2
  \lambda_V^2} \nonumber\\
Z(\tau)=\frac{c_2 e^{\tau /2}}{\sqrt{e^{2 c_1 \lambda_V^2+\tau }-2
\lambda_V^2
   e^{3 c_1+\frac{\lambda_W \tau }{\lambda_V}}}},	
\end{align}
where $c_1$ and $c_2$ are integration constants. 
In this case, the system \eqref{ODEs1} admits two classes of critical
points: the point $P_3$ for which the stability conditions reduce to
$\frac{\lambda_W}{\lambda_V}<1, \lambda_V^2\geq \frac{3}{2}$ and the
$Z$-axis which is a curve of equilibrium points containing the points $P_1$
and $P_2$. The center direction of the curve is tangent to the   
$Z$-axis, and therefore this curve of critical points is normally
hyperbolic
\cite{Aulbach1984a} (a set of non-isolated singular points is
called normally
hyperbolic if the only eigenvalues with zero real parts are those
whose corresponding eigenvectors are tangent to the set), and
since by definition any point on a set of non-isolated singular
points will have at least one eigenvalue which is zero, all points
in the set are   non-hyperbolic.  The stability of a set which is
normally hyperbolic can  be completely classified   by considering the
signs of the eigenvalues in the
remaining directions \cite{Aulbach1984a}. In conclusion, in the special
case $\gamma=1$, the curve of critical points that contains $P_1$ and
$P_2$ is stable for $\frac{\lambda_W}{\lambda_V}>1$.

Finally, lets us comment on the asymptotic behavior of $P_3$.
From  the constraint equation (\ref{constraint}) it follows that 
\begin{equation}
\frac{d\psi}{d \tau}=\frac{f_1(a_{ref} e^{-\tau})}{\lambda_V f_2(a_{ref}
e^{-\tau})},
\end{equation} 
which has the solution
\begin{equation}
\lambda_V (\psi-\psi_0)=\int_{0}^\tau \frac{ f_1(a_{ref} e^{-\eta})}{
f_1(a_{ref} e^{-\eta})} d \eta,
\end{equation} where the current scale factor is set to 1 and $\psi_0$
denotes the current $\psi$-value. 
Hence,  
\begin{align}
V\propto e^{-\lambda_V (\psi-\psi_0)}
 =\frac{(1-a_{ref}) \left[-3 \alpha_3 (a_{ref}-1)+\alpha_4
(a_{ref}-1)^2+3\right]}{\left(e^{\tau }-a_{ref}\right) \left[e^{2 \tau } (3
   \alpha_3+\alpha_4+3)+\alpha_4 a_{ref}^2-a_{ref} e^{\tau } (3 \alpha_3+2
\alpha_4)\right]},
\end{align}
while
\begin{align}
W\propto  e^{-\lambda_W (\psi-\psi_0)}=\left\{\frac{(1-a_{ref}) \left[-3
\alpha_3 (a_{ref}-1)+\alpha_4 (a_{ref}-1)^2+3\right]}{\left(e^{\tau
}-a_{ref}\right) \left[e^{2 \tau } (3
   \alpha_3+\alpha_4+3)+\alpha_4 a_{ref}^2-a_{ref} e^{\tau } (3 \alpha_3+2
\alpha_4)\right]}\right\}^{\frac{\lambda_W}{\lambda_V}}.
\end{align}
Therefore, since  $a_{ref}\lesssim10^{-9}$, if
$\frac{\lambda_W}{\lambda_W}>0$ both $V$ and $W$ tend to zero as
$\tau\rightarrow +\infty$.  However, for
$\frac{\lambda_W}{\lambda_W}<0$, $V$ tends to zero but $W$ tends to
infinity as $\tau\rightarrow \infty$, and since $Y_c\neq 0$   we
deduce that $H\rightarrow \infty$ as $\tau\rightarrow \infty$. This is a
Big-Rip-type behavior, however it is realized at infinity and not at a
finite scale factor
\cite{Sami:2003xv,Nojiri:2005sx,Copeland:2006wr,Briscese:2006xu,
Bamba:2008ut,Capozziello:2009hc,Saridakis:2009jq}.

  \subsection{Open universe}
\label{App2}

Let us discuss   the critical points of the autonomous system
\eqref{ODEs2} and their stability conditions.  From the last
equation of \eqref{ODEs2} it follows that either $q=0$ or $\Omega_k=0$, and
therefore we  can simplify the investigation and examine these two cases
separately.

Note that the variable choice
(\ref{auxiliary2}) allows for an easy, partial, classification of
expanding and contracting solutions. In particular, solutions with  
$\Omega_k=k/(aH)>0$ correspond to $H>0$ and thus to expansion, while those
with $\Omega_k<0$ correspond to $H<0$ and therefore to contraction
($k=\sqrt{|K|}$ throughout this work). That is why points with
$\Omega_k>0$ are denoted with the subscript ``+'', while those with
$\Omega_k<0$ are denoted with the subscript ``-''. However, this is only 
a partial classification, since it cannot work  for solutions with
$\Omega_k=0$, which can be either expanding or contracting.
Finally, we mention that although our model admits expanding and 
contracting solutions, since the sign of $\Omega_k$ is invariant, 
there is no transition from contracting to expanding solutions or vice
versa.  Nevertheless, there could still exist  at the edge of the phase
space, and in such a case they could  be revealed only through
application of Poincar\'e central projection method
\cite{PoincareProj,Leon2011,Leon:2012mt}. This
analysis lies beyond the scope of the present work and it is left for
future investigation.

\subsubsection*{Analysis in the invariant set $\Omega_k=0$}

In this case, from the first equation of  \eqref{ODEs2} it follows that
$X=0$.
Thus, the curvatureless equilibrium solutions must satisfy 
\begin{eqnarray}
&&Y \left\{\sqrt{6} U^2 \left[(\gamma -2) \lambda_V-\lambda_W\right]+2 U
(\lambda_V
\lambda_W+3)\right. \nonumber \\ 
&& \ \ \ \ \ \ \ \ \ \ \ \ \   
\left.
-\sqrt{6} \left\{\lambda_W Y+\lambda_V \left[\gamma -\gamma  Y+(\gamma -1)
Z
(4\alpha_3+\alpha_4+6)\right]\right\}\right\}=0,\\
&&Z \left\{-\sqrt{6} (\gamma -3) \lambda_V U^2-2 \left(\lambda_V^2+3\right)
U \right. \nonumber   \\ 
&& \ \ \ \ \ \ \ \ \ \ \ \ \   
\left.
+\sqrt{6} \left\{\lambda_W Y+\lambda_V \left[\gamma -\gamma  Y+(\gamma -1)
Z
(4\alpha_3+\alpha_4+6)\right]\right\}\right\}=0,\\
	&& U\neq \frac{\sqrt{6}}{3} \lambda_V.
\end{eqnarray}
Note that in this case the evolution equation for $U$ reduces to $U'=0$,
which implies that in the former expressions $U$ behaves as a
parameter (a constant). 

Thus, in the case of $\Omega_k=0$ we have the following curves of critical
points:
\begin{itemize}
\item Curve $Q_1$: $X_{c 1}=0, Y_{c 1}= \frac{\lambda_V \left[\gamma
+(\gamma -1) Z_c (4
\alpha_3+\alpha_4+6)\right]}{\gamma  \lambda_V-\lambda_W}, Z_{c 1}=Z_c,
U_{c
1}=0, {\Omega_k}_{c 1}=0$,
with eigenvalues $$\left\{-1,-1,0,0,-3 \gamma\right\}.$$

\item Curve $Q_2$: $X_{c 2}=0, Y_{c 2}=0, Z_{c 2}=Z_c,$ \\ 
$U_{c2}=-\frac{\lambda_V^2+\sqrt{\lambda_V^4+6 \lambda_V^2 \left[(\gamma -3)
\gamma
+(\gamma -3) (\gamma -1) Z_c (4
\alpha_3+\alpha_4+6)+1\right]+9}+3}{\sqrt{6}
(\gamma -3) \lambda_V}$, ${\Omega_k}_{c 2}=0$,
with eigenvalues

$\Big\{0,-1,-\frac{2 \gamma +\lambda_V^2+\sqrt{\lambda_V^4+6 \lambda_V^2
\left[(\gamma
-3) \gamma +(\gamma -3) (\gamma -1) Z_c (4
\alpha_3+\alpha_4+6)+1\right]+ 9}-3}{2
(\gamma -3)},$

$\frac{6 (\gamma -3) (\gamma -1) \lambda_V^2 Z_c (4
\alpha_3+\alpha_4+6)}{(2 \gamma -5)
   \lambda_V^2+\sqrt{\lambda_V^4+6 \lambda_V^2 \left[(\gamma -3) \gamma
+(\gamma
-3) (\gamma -1) Z_c (4 \alpha_3+\alpha_4+6)+1\right]+9}+3},$  

	$-\frac{(\lambda_V-\lambda_W) \left(\lambda_V^2+\sqrt{\lambda_V^4+6
\lambda_V^2 \left[(\gamma -3) \gamma +(\gamma -3)
   (\gamma -1) Z_c (4 \alpha_3+\alpha_4+6)+1\right]+9}+3\right)}{(\gamma
-3)
\lambda_V}\Big\}.$ 

\item Curve $Q_3$: $X_{c 3}=0, Y_{c 3}=0, Z_{c 3}=Z_c,$ \\
$U_{c3}=\frac{-\lambda_V^2+\sqrt{\lambda_V^4+6 \lambda_V^2 \left[(\gamma -3)
\gamma
+(\gamma -3) (\gamma -1) Z_c (4
\alpha_3+\alpha_4+6)+1\right]+9}-3}{\sqrt{6}
(\gamma -3) \lambda_V}$, ${\Omega_k}_{c 3}=0$,
with eigenvalues

$\Big\{0,-1,\frac{-2 \gamma -\lambda_V^2+\sqrt{\lambda_V^4+6 \lambda_V^2
\left[(\gamma
-3) \gamma +(\gamma -3) (\gamma -1) \text{Zc} (4
\alpha_3+\alpha_4+6)+1\right]+9}+3}{2 (\gamma -3)},$ 
	
	$\frac{6 (\gamma -3) (\gamma -1) \lambda_V^2 Z_c (4
\alpha_3+\alpha_4+6)}{(2 \gamma -5)
   \lambda_V^2-\sqrt{\lambda_V^4+6 \lambda_V^2 \left[(\gamma -3) \gamma
+(\gamma
-3) (\gamma -1) Z_c (4 \alpha_3+\alpha_4+6)+1\right]+9}+3},$   
	
	$\frac{(\lambda_V-\lambda_W) \left(-\lambda_V^2+\sqrt{\lambda_V^4+6
\lambda_V^2 \left[(\gamma -3) \gamma +(\gamma -3)
   (\gamma -1) \text{Zc} (4
\alpha_3+\alpha_4+6)+1\right]+9}-3\right)}{(\gamma
-3) \lambda_V}\Big\}.$
	
\item Curve $Q_4$:  $X_{c 4}=0, Y_{c 4}=0, Z_{c 4}=0, U_{c
4}=U_c,{\Omega_k}_{c 4}=0$,
with eigenvalues	

$\Big\{0,-1,\frac{\sqrt{6} (2-3 \gamma ) \lambda_V+3 \sqrt{6} (\gamma -2)
\lambda_V U_c^2+12 U_c}{6 U_c-2 \sqrt{6} \lambda_V},\frac{3 \left[-\sqrt{6}
\gamma  \lambda_V+\sqrt{6} (\gamma -3) \lambda_V U_c^2+2
\left(\lambda_V^2+3\right) U_c\right]}{3
   U_c-\sqrt{6} \lambda_V},$   
	
	$\frac{3 \left\{-\sqrt{6} \gamma  \lambda_V+\sqrt{6} U_c^2
\left[(\gamma
-2) \lambda_V-\lambda_W\right]+2 U_c
   (\lambda_V \lambda_W+3)\right\}}{3 U_c-\sqrt{6} \lambda_V}\Big\}.$
	
\item Curve $Q_5$: $X_{c 5}=0, Y_{c 5}=1-U_c^2+\frac{\left(\sqrt{6}
\lambda_V
U_c-3\right) \left(\sqrt{6} U_c-\lambda_W\right)}{3 \gamma  \lambda_V-3
\lambda_W}, Z_{c 5}=0, U_{c 5}=U_c,  {\Omega_k}_{c 5}=0.$ In order to
determine the stability of this curve of critical points we  need to resort
to numerical inspection. 
\end{itemize}

The examination of the sign of the above eigenvalues is straightforward
for the general case $\gamma\neq1$, however in the special case
$\gamma=1$, which is the most interesting   in physical terms since it
corresponds to dust matter, some eigenvalues become zero and thus the
corresponding curves of critical points become non-hyperbolic. In this
case if the remaining eigenvalues have different sign then the  curve of
critical points
behaves like saddle, while if they are of the same sign then the
non-hyperbolic curve of critical points has a stable or unstable manifold
of smaller dimensionality (in principle one must apply the center manifold
theorem \cite{Aulbach1984a}). The  curves of
critical points $Q_1$-$Q_5$ for the special case
$\gamma=1$  are summarized in Table \ref{crit01open1}, while their
stability conditions are displayed in Table \ref{crit01open2}.

\subsubsection*{Analysis in the invariant set $q=0$}

In the  case $q=0$, from \eqref{ODEs2} we deduce that the equilibrium
solutions must satisfy one of the following three possibilities:
\begin{itemize} 
\item $Y_c\neq 0, Z_c=0, U_c=\frac{\sqrt{6}}{3\lambda_W}$,
\item $Y_c= 0, Z_c\neq 0, U_c=\frac{\sqrt{6}}{3\lambda_V}$,
\item $Y_c=0,Z_c=0$.
\end{itemize}
In the first case,   substituting the values of $Z_c=0,
U_c=\frac{\sqrt{6}}{3\lambda_W}$ into the fourth equation of
\eqref{ODEs2} we conclude that the
equilibrium solution satisfies $Y_c=\frac{4}{3\lambda_W^2}$. Inserting this
 into the expression for $q$ we obtain the additional constraint
$-\frac{(3 \gamma -2) \left[\lambda_W^2
\left(\Omega_k^2-1\right)+2\right]}{2 \lambda_W^2}=0$, which leads to
$\Omega_{k c}=\pm \sqrt{1-\frac{2}{\lambda_W^2}}$ (corresponding to
expanding and contracting universe respectively). Finally, inserting these
expressions in the relation for $\dot b$  (\ref{eq42bb}) we find that
$\Omega_k\dot{b}=X$, and thus the first equation of \eqref{ODEs2} is
satisfied identically, irrespectively the
value of $X$. In summary, in this case we obtain two curves of
critical points, namely
$$Q_6^+: \ X_{c 6}^+=X_c, Y_{c 6}^+=\frac{4}{3\lambda_W^2}, Z_{c 6}^+=0, 
U_{c
6}^+=\frac{\sqrt{6}}{3\lambda_W}, 
\Omega_{kc
6}^+=\sqrt{1-\frac{2}{\lambda_W^2}},$$ and 
$$Q_6^-: \ X_{c 6}^-=X_c, Y_{c 6}^-=\frac{4}{3\lambda_W^2}, Z_{c 6}^-=0, 
U_{c
6}^-=\frac{\sqrt{6}}{3\lambda_W}, \Omega_{kc
6}^-=-\sqrt{1-\frac{2}{\lambda_W^2}}.$$

In the second case, the system admits two critical points, namely
$$Q_7^+: \ X_{c 7}^+=0, Y_{c 7}^+=0, Z_{c 7}^+=-\frac{4}{3 \lambda_V^2 (4
\alpha_3+\alpha_4+6)}, U_{c 7}^+=\frac{\sqrt{6}}{3\lambda_V},
\Omega_{kc
7}^+=\sqrt{1-\frac{2}{\lambda_V^2}}$$
and
$$Q_7^-: \  X_{c 7}^-=0, Y_{c 7}^-=0, Z_{c 7}^-=-\frac{4}{3 \lambda_V^2 (4
\alpha_3+\alpha_4+6)}, U_{c 7}^-=\frac{\sqrt{6}}{3\lambda_V},
\Omega_{kc
7}^-=-\sqrt{1-\frac{2}{\lambda_V^2}}.$$

Finally, in  the third case, from the fourth equation of \eqref{ODEs2} it
follows that $U_c=0$. Thus, substituting $Y_c=0,Z_c=0,U_c=0$ in the rest
of the equations, and assuming that
$\gamma\neq \frac{2}{3}$, we obtain $\Omega_{k c}=\pm 1$. Therefore, for
$\Omega_{k c}=+1$ the first equation of \eqref{ODEs2}   gives
\begin{eqnarray}
&&Q_{8}: \  X_{c 8}=1, Y_{c 8}=0, Z_{c 8}=0, U_{c
8}=0, \Omega_{k c8}=1\nonumber\\
&&Q_{9}: \  X_{c 9}=\frac{-\sqrt{4 \alpha_3^2-6 \alpha_4}+2
\alpha_3+\alpha_4}{\alpha_4}, Y_{c 9}=0, Z_{c 9}=0, U_{c
9}=0, \Omega_{k c9}=1\nonumber\\
&&Q_{10}: \  X_{c 10}=\frac{\sqrt{4 \alpha_3^2-6 \alpha_4}+2
\alpha_3+\alpha_4}{\alpha_4}, Y_{c 10}=0, Z_{c 10}=0, U_{c
10}=0, \Omega_{k c10}=1,\nonumber
\end{eqnarray}
while for  $\Omega_{k c}=-1$, we
obtain that $X=\tilde{X}_{c 11}$, where $\tilde{X}_{c 11}$
is the unique real solution of the equation  $-2 \alpha_3
\left(X^2+X-2\right)+\alpha_4 (X+1) (X-1)^2+6=0.$

Lets us now examine the  
  eigenvalues associated to the critical points or curves
$Q_6^\pm$-$Q_{11}$.
The eigenvalues of $Q_6^\pm$  are 
$\left\{0,2-3 \gamma ,2-\frac{2 \lambda_V}{\lambda_W},-\frac{\sqrt{8
\lambda_W^2-3 \lambda_W^4}}{\lambda_W^2}-1,\frac{\sqrt{8
   \lambda_W^2-3 \lambda_W^4}}{\lambda_W^2}-1\right\}$. The eigenvalues of
$Q_7^\pm$ (for $\gamma=1$)    are
 $\left\{0,-1 ,2-\frac{2 \lambda_V}{\lambda_W},\Delta_1(\alpha_3,\alpha_4,
\lambda_V,\lambda_W),\Delta_2(\alpha_3,\alpha_4,
\lambda_V,\lambda_W)\right\}$, where $\Delta_{1,2}(\alpha_3,\alpha_4,
\lambda_V,\lambda_W)$ are complicated functions of their arguments
 that
can be obtained explicitly only by numerical elaboration.  	
The eigenvalues associated to $Q_{8}$ are $\{0,2,2,-2,2-3\gamma\}$.
The eigenvalues associated to $Q_{9}, Q_{10}$ are
$\{-2,2,2,0,\Delta_3(\gamma,\alpha_3,\alpha_4)\}$ and finally for $Q_{11}$
they are $\{-2,2,2,0,\Delta_4(\gamma,\alpha_3,\alpha_4)\}$, where
$\Delta_{3,4}(\gamma,\alpha_3,\alpha_4)$ are   complicated expressions
of their arguments. Thus,   $Q_{8}$-$Q_{11}$ are always saddle since
at least two eigenvalues
have different signs. 

The  individual critical points $Q_7^+$-$Q_{10}$ and the  curves of
critical points $Q_6^\pm$ and $Q_{11}$, for
the special case $\gamma=1$,  are summarized in Table \ref{crit01open1},
while their
stability conditions are displayed in Table \ref{crit01open2}.

\subsubsection*{Quintessence-like solutions}

We close this Appendix by mentioning that the curve of critical points
$Q_5$ analyzed above includes many interesting cosmological solutions, and
in particular the points of standard quintessence 
\cite{Copeland:1997et,Copeland:2009be}. Focusing for simplicity on the case
$\gamma=1$, these points were presented in Table \ref{quintessence}.
However, the stability conditions are different than the usual conditions
in \cite{Copeland:1997et,Copeland:2009be} due to the presence of extra
phase-space directions, namely those of curvature and graviton mass. 
  \begin{table*}[[ht]
\begin{center}
\begin{tabular}{|c|c|c|}
\hline
&&   \\
 Cr. P. & Eingenvalues & Stability \\
\hline \hline
$Q_{12}$&   $2,-1,0,\sqrt{6} \lambda_V+6,\sqrt{6} \lambda_W+6$ & saddle
point
\\[0.2cm]\hline
$Q_{13}$& $2,-1,0,6-\sqrt{6} \lambda_V,6-\sqrt{6} \lambda_W$ & saddle point
\\[0.2cm]\hline
$Q_{14}$ & $-1,0,\frac{\left(\lambda_W^2-6\right) (\lambda_V-\lambda_W)}{2
\lambda_V-\lambda_W},\frac{1}{2} \left(\lambda_W^2-2\right),\lambda_W
(\lambda_W-\lambda_V)$ & stable
node for  \\[0.2cm]
&  &  $-\sqrt{2}<\lambda_W<0,\lambda_V<\lambda_W$ or  \\[0.2cm]
&  &  $0<\lambda_W<\sqrt{2}, \lambda_V>\lambda_W$  \\[0.2cm]
&  &  saddle point otherwise \\[0.2cm]
 \hline
$Q_{15}$& $-1,\frac{1}{2},0,-\frac{3
(\lambda_V-\lambda_W)}{\lambda_W},-\frac{9 (\lambda_V-\lambda_W)}{\lambda_W
(2
   \lambda_V \lambda_W-3)}$  &  saddle point  \\[0.2cm]
 \hline
$Q_{16}$& $3,3,-1,\frac{1}{2},0$  &  saddle point  \\[0.2cm]
 \hline
$Q_{17}$& $2,2,-1,0,2$  &  saddle point  \\[0.2cm]
 \hline
$Q_{18}$& $2,2,-1,-1,-4$  &  saddle point  \\[0.2cm]
 \hline
\end{tabular}
\end{center}
\caption[crit]{\label{quintessenceb} The eigenvalues of   matrix
${\bf {Q}}$ of the perturbation equations of the autonomous system
(\ref{ODEs2}), and the corresponding stability conditions, for the
quintessence-like solutions presented in Table \ref{quintessence}.}
\end{table*}

In particular, for the critical points $Q_{12}$ to $Q_{18}$ of Table
\ref{quintessence}, the coefficients of the perturbation equations form a
$5\times5$ matrix ${\bf {Q}}$, that allows for an easy calculation of its
eigenvalues. The corresponding eigenvalues and the stability conditions for
each critical point are   displayed in Table \ref{quintessenceb}.
Finally,  some of these points possess one zero eigenvalue and  are
thus  non-hyperbolic. In the case of normally-hyperbolic curves of critical
points  (that is the only eigenvalues with zero real parts are those
whose corresponding eigenvectors are tangent to the set) the stability is
extracted considering the signs
of  the rest eigenvalues \cite{Aulbach1984a}. For isolated non-hyperbolic
critical points we can determine the dimensionality of their stable
manifold using the linearization technique \cite{Aulbach1984a}.

 \section{Stability when $b(t)$ is imposed at will}

\subsection{Flat universe}
\label{App3}

For the critical points $(x_c,y_c,u_c,v_c)$ of the autonomous system
system  (\ref{autonomous1})-(\ref{autonomous4}), the coefficients of the
perturbation
equations form a $4\times4$ matrix ${\bf {Q}}$, which using
the specific critical points presented in Table \ref{crit}  acquires a
simple form that allows for an easy calculation of
its eigenvalues.   The corresponding
eigenvalues  and the stability
conditions
for each critical point are presented in Table
\ref{Tab3b}. 
\begin{table*}[ht]
\begin{center}
\begin{tabular}{|c|c|c|}
\hline
&&   \\
 Cr. P.& Eigenvalues &
Stability \\
\hline \hline
$R_1$& $-\frac{3}{2},\, \frac{3}{2},\, -1,\, 0$& non-hyperbolic (behaves
as saddle point) \\
\hline
 $R_2$& $3,\, 3,\, -1,\, 3-\sqrt{\frac{3}{2}}\lambda_W$ &     saddle point
\\
\hline
 $R_3$& $3,\, 3,\, -1,\, 3+\sqrt{\frac{3}{2}}\lambda_W$ &     saddle point 
\\
\hline
$R_4^\pm$ &  $-1,-3+\frac{\lambda_W^2}{2},-3+{\lambda_W^2},
-3+{\lambda_W^2}$ &
stable
node for $\lambda_W^2<3$\\
& & saddle point for
 $3<\lambda_W^2<6$ \\
\hline
 $R_5^\pm$& $0,-1, \alpha^-(\lambda_W), \alpha^+(\lambda_W) $ &
non-hyperbolic
\\
& & 3D stable manifold for \\
& &  $3<\lambda_W^2<\frac{24}{7}$ or
\\
& &  $\lambda_W^2>\frac{24}{7}$\\
\hline
$R_6^\pm$& $-3,-3,-3,-1$ & stable node \\
\hline
$R_7$& $-1,0,\beta^-(\lambda_W,y_c),\beta^+(\lambda_W,y_c)$ & normally
hyperbolic (behaves as stable) \\
\hline
$R_8$& $-3,-1, 0, \frac{3}{2}-\sqrt{\frac{3}{2}}\lambda_W x_c$ & stable for
$x_c \lambda_W>\sqrt{\frac{3}{2}};$  
\\
& & saddle point otherwise. \\
\hline
\end{tabular}
\end{center}
\caption[crit]{\label{Tab3b}
The eigenvalues of   matrix
${\bf {Q}}$ of the perturbation equations of the autonomous system
(\ref{autonomous1})-(\ref{autonomous4}), calculated at the critical points
  presented in Table
\ref{crit}, and their stability conditions. We have introduced the
notations $\alpha^\pm(\lambda_W)=\frac{3}{4} \left(-1\pm\frac{\sqrt{24
\lambda_W ^2-7 \lambda_W ^4}}{\lambda_W ^2}\right),$  and
$\beta^\pm(\lambda_W,
y_c)=-\frac{1}{2} \left[3-\lambda_W ^2 y_c^2\pm\sqrt{\lambda_W^4 y_c^4-18
   \left(\lambda_W ^2-2\right) y_c^2+9}\right].$}
\end{table*}
We mention that point $R_1$ is non-hyperbolic, but since it
has eigenvalues with different sign, and using the   center manifold
analysis \cite{Aulbach1984a}, we can straightforwardly show that it behaves
as
saddle point. Moreover, the  non-hyperbolic curve of critical 
points $R_7$ has a central direction normal to the set and therefore it
behaves as stable. Finally, note that although $R_6^\pm$ at first sight
seems
to be the $\lambda_W\rightarrow0$ limit of $R_4^\pm$   this
is not the case since the complete equations are different.

  \subsection{Open universe}
\label{App4}

For the critical points $(x_c,y_c,u_c,v_c,\Omega_{kc})$ of the autonomous
system system  (\ref{autonomousb1})-(\ref{autonomousb5}), the coefficients
of the perturbation equations form a $5\times5$ matrix ${\bf {Q}}$,
which using the specific critical points presented in Table \ref{critb} 
acquires a simple form that allows for an easy
calculation of its eigenvalues. The corresponding
eigenvalues for each critical point are presented in Table
\ref{critbeigen}. Finally, note that although $S_6^\pm$ at first sight
seems
to be the $\lambda_W\rightarrow0$ limit of $S_4^\pm$   this
is not the case since the complete equations are different.

 \begin{table*}[[ht]
\begin{center}
\begin{tabular}{|c|c|c|}
\hline
&&    \\
 Cr. P.& Eigenvalues &
Stability  \\
\hline \hline
$S_1$& $-\frac{3}{2},\, \frac{3}{2},\, -1,\, 3,\, \frac{1}{2}$& saddle
point
\\
\hline
 $S_2$& $6,\,2,\, 3,\, -1,\, 3-\sqrt{\frac{3}{2}}\lambda_W$ &     saddle
point
\\
\hline
 $S_3$& $6,\,2,\, 3,\, -1,\, 3+\sqrt{\frac{3}{2}}\lambda_W$ &     saddle
point 
\\
\hline
$S_4^\pm$ &  $\lambda_W^2,\,-1,-3+\frac{\lambda_W^2}{2},-3+{\lambda_W^2},
-1+\frac{\lambda_W^2}{2}$ & saddle point\\
\hline
 $S_5^\pm$& $-1, \alpha^-(\lambda_W), \alpha^+(\lambda_W), 3,\frac{1}{2} $
&
saddle point \\
\hline
$S_6^\pm$& $0, -3,-3,-3,-1$ & non-hyperbolic (4D stable manifold) \\
\hline
$S_7^\pm$& $-2,2,-1,-1,1$ & saddle point\\
\hline
$S_8^\pm$&
$-1,-1,2,-1+\sqrt{-3+\frac{8}{\lambda_W^2}},-1-\sqrt{-3+\frac{8}{
\lambda_W^2}}$
& saddle point\\
\hline
\end{tabular}
\end{center}
\caption[crit]{\label{critbeigen}
The eigenvalues of   matrix
${\bf {Q}}$ of the perturbation equations of the autonomous system
(\ref{autonomousb1})-(\ref{autonomousb5}), calculated at the critical
points
  presented in Table
\ref{critb}, and their stability conditions. We have introduced the
notations  $\alpha^\pm(\lambda_W)=\frac{3}{4} \left(-1\pm\frac{\sqrt{24
\lambda_W ^2-7 \lambda_W ^4}}{\lambda_W ^2}\right).$ }
\end{table*}

In order to examine the corresponding stability
conditions we have to examine the sign of these eigenvalues. An
interesting observation  from \eqref{autonomousb4} is that the
sign of $v$ (which according to (\ref{auxiliaryb}) is the auxiliary
variable proportional to the graviton mass square) is invariant. Therefore,
$v$ remains zero if initially it is zero, and in this case we
can examine the system in the invariant set $v=0$.
In this case the possible late-time solutions are either   $S_4^\pm$
provided $\lambda_W^2<2$ or either $S_8^\pm$ 
for $\lambda_W^2>2$. In the particular case of $2<\lambda_W^2\leq
\frac{8}{3},$
the points  $S_8^\pm$ are spiral attractors in a 2D sub-manifold
(two negative
real eigenvalues and two complex conjugated eigenvalues with negative real
part). Finally, points $S_6^\pm$ are non-hyperbolic, with a 4D stable
manifold.

However, in the case where $v\neq0$ only points $S_6^\pm$ behave as
stable,
since all the other become saddle points. In particular, introducing the
local coordinates
$
\left\{x-x_c,y-y_c,u, v,\Omega_k\right\}=\epsilon
\left\{\widetilde{x},\widetilde{y},\widetilde{u},
\widetilde{v},\widetilde{\Omega_k}\right\}+{\cal O}(\epsilon)^2$
where $\epsilon$ is a constant satisfying $\epsilon\ll 1,$ we deduce that 
\begin{align}
&\widetilde{v}'=3 \widetilde{v} \left(x_c^2-y_c^2+1\right)+ \text{h.o.t}
\nonumber\\
&\widetilde{\Omega_k}'=\frac{1}{2} \widetilde{\Omega_k}
\left(3
x_c^2-3 y_c^2+1\right)-\frac{1}{2} \widetilde{v} \beta  \left[\beta ^2
(\alpha_3+\alpha_4)-2 \beta  (2 \alpha_3+\alpha_4+1)+3
\alpha_3+\alpha_4+3\right]+ \text{h.o.t},
\end{align} 
where $x_c$ and $y_c$ are  the coordinates of the critical
points $S_1$ to $S_5$ and   $\text{h.o.t}$ denoting ``higher order
terms''. These equations admit the general solutions
\begin{align}
&\widetilde{v}= c_1 e^{3 \tau  \left(x_c^2-y_c^2+1\right)},\nonumber\\
&\widetilde{\Omega_k}= \frac{c_1 \beta  \left[e^{\frac{1}{2} \tau  \left(3
x_c^2-3
   y_c^2+1\right)}-e^{3 \tau  \left(x_c^2-y_c^2+1\right)}\right]
\left(\alpha_3 \beta ^2-4 \alpha_3 \beta +3 \alpha_3+\alpha_4 \beta ^2-2
\alpha_4 \beta +\alpha_4-2 \beta +3\right)}{3 x_c^2-3
y_c^2+5}\nonumber\\&
  \ \ \ \ \ \ \ +c_2
e^{\frac{1}{2} \tau  \left(3
   x_c^2-3 y_c^2+1\right)},
\end{align}
which implies that the system is unstable in $v$ and $\Omega_k$ directions.

In the special case of   point $S_6^+,$ using a similar approach we extract
that the perturbations $\widetilde{v}$ and
$\widetilde{\Omega_k}$ satisfy the equations
\begin{equation}\label{pertsS6}
\widetilde{v}'=-\frac{\vartheta \widetilde{v}^2}{\widetilde{\Omega_k}},\; 
\widetilde{\Omega_k}'=\frac{1}{2} (-\widetilde{v} \vartheta -2
\widetilde{\Omega_k}),
\end{equation}
where $\vartheta =\beta  \left[\beta ^2 (\alpha_3+\alpha_4)-2 \beta  (2
\alpha_3+\alpha_4+1)+3 \alpha_3+\alpha_4+3\right]$.   
The system \eqref{pertsS6} admits two general solutions 
\begin{equation}
 \widetilde{v}=\frac{4 e^{2 c_2}}{\left(e^{\tau }-e^{c_2} c_1 \vartheta
\right){}^2},\; \widetilde{\Omega_k}=\frac{2 \vartheta  e^{2 c_2-\tau
}}{e^{\tau }-e^{c_2} c_1 \vartheta }
\end{equation}
and 
\begin{equation}
 \widetilde{v}=\frac{4 e^{2 c_2}}{\left(e^{c_2} c_1 \vartheta +e^{\tau
}\right){}^2},\; \widetilde{\Omega_k}=\frac{2 e^{2 c_2} \vartheta
}{\left(e^{c_2} c_1 \vartheta +e^{\tau }\right) \left(2 e^{c_2} c_1
   \vartheta +e^{\tau }\right)},
\end{equation}
where $c_1$ and $c_2$ are integration constants.
In both cases the $v$-perturbations and $\Omega_k$-perturbations  decay  to
zero in the limit $\tau\rightarrow +\infty$, and thus points $S_6^\pm$ are
stable.

The stability conditions for the critical points $S_1$-$S_8^\pm$ are
summarized in  Table \ref{critbeigen}.

\end{appendix}

\end{document}